\documentclass[]{aa} 

\usepackage{graphicx}
\usepackage{txfonts}
\usepackage{amsmath,amstext}

\usepackage{longtable}

\usepackage{multicol}
\usepackage{multirow}

\usepackage{textgreek}

\usepackage[]{hyperref}
\hypersetup{
colorlinks=true,
linkcolor=blue,
urlcolor=black,
citecolor=black
}
\makeatletter
\renewcommand*\aa@pageof{, page \thepage{} of \pageref*{LastPage}}
\makeatother

\usepackage{epsf}
\usepackage{rotating}

\usepackage{color}

\newcommand{\teff}{\mbox{$T_{\rm eff}$}}
\newcommand{\logg}{\mbox{$\log{g}$}}
\newcommand{\feh}{\mbox{$[\mathrm{Fe/H}]$}}
\newcommand{\afe}{\mbox{$[\mathrm{\alpha/Fe}]$}}
\newcommand{\vsini}{\mbox{$v \sin i_*$}}
\newcommand{\mictrb}{\mbox{$\xi_{\rm t}$}}
\newcommand{\mactrb}{\mbox{$v_{\rm mac}$}}

\newcommand{\kms}{\mbox{km\,s$^{-1}$}}
\newcommand{\ms}{\mbox{m\,s$^{-1}$}}

\newcommand{\pdf}{\textsc{pdf}}
\newcommand{\prior}{\textit{prior}}
\newcommand{\post}{\textit{posterior}}

\renewcommand{\eqref}[1]{\ref{eq:#1}}

\newcommand{\sectionname}{Section}
\newcommand{\sectref}[1]{\ref{sect:#1}}
\newcommand{\Sect}[1]{\sectionname~\sectref{#1}}
\newcommand{\sect}[1]{\Sect{#1}}
\newcommand{\sectlabel}[1]{\label{sect:#1}}

\newcommand{\tabname}{Table}
\newcommand{\tabref}[1]{\ref{tab:#1}}
\newcommand{\Tab}[1]{\tabname~\tabref{#1}}
\newcommand{\tab}[1]{\Tab{#1}}
\newcommand{\tablabel}[1]{\label{tab:#1}}
\newcommand{\tabalt}[1]{\tabref{#1}}
\newcommand{\figname}{Fig}
\newcommand{\figref}[1]{\ref{fig:#1}}
\newcommand{\Fig}[1]{\figname~\figref{#1}}
\newcommand{\fig}[1]{\Fig{#1}}
\newcommand{\figlabel}[1]{\label{fig:#1}}

\begin{document}
      
\title{The discovery of WASP-151b, WASP-153b, WASP-156b: Insights on giant planet migration and the upper boundary of the Neptunian desert}

\author{Olivier. D. S. Demangeon\inst{\ref{Porto}, \thanks{\email{olivier.demangeon@astro.up.pt}}} \and
F. Faedi\inst{\ref{Warwick}, \ref{INAF-Catania}} \and 
G. H\'{e}brard\inst{\ref{IAP}, \ref{OHP}} \and 
D. J. A. Brown\inst{\ref{Warwick}} \and 
S. C. C. Barros\inst{\ref{Porto}} \and
A. P. Doyle\inst{\ref{Warwick}} \and
P. F. L. Maxted\inst{\ref{Keele}} \and
A. Collier Cameron\inst{\ref{StAndrews}} \and
K. L. Hay\inst{\ref{StAndrews}} \and
J. Alikakos\inst{\ref{Greece}} \and
D. R. Anderson\inst{\ref{Keele}} \and
D. J. Armstrong\inst{\ref{Warwick}, \ref{Belfast}} \and
P. Boumis\inst{\ref{Greece}} \and
A. S. Bonomo\inst{\ref{INAF}} \and
F. Bouchy\inst{\ref{Geneve}} \and
C. A. Haswell\inst{\ref{OpenUniv}} \and
C. Hellier\inst{\ref{Keele}} \and
F. Kiefer\inst{\ref{IAP}} \and
K. W. F. Lam\inst{\ref{Warwick}} \and
L. Mancini\inst{\ref{Rome}, \ref{MaxPlanck}, \ref{INAF}} \and
J. McCormac\inst{\ref{Warwick}} \and
A. J. Norton\inst{\ref{OpenUniv}} \and
H. P. Osborn\inst{\ref{Warwick}} \and
E. Palle\inst{\ref{IAC}, \ref{ULL}} \and
F. Pepe\inst{\ref{Geneve}} \and
D. L. Pollacco\inst{\ref{Warwick}} \and
J. Prieto-Arranz\inst{\ref{IAC}, \ref{ULL}} \and
D. Queloz\inst{\ref{Cambridge}, \ref{Geneve}} \and
D. S\'egransan\inst{\ref{Geneve}} \and
B. Smalley\inst{\ref{Keele}} \and
A. H. M. J. Triaud\inst{\ref{Cambridge-IA}, \ref{Birmingham}} \and
S. Udry\inst{\ref{Geneve}} \and
R. West\inst{\ref{Warwick}} \and
P.J. Wheatley\inst{\ref{Warwick}}
}

\institute{Instituto de Astrof\'isica e Ci\^{e}ncias do Espa\c co, Universidade do Porto, CAUP, Rua das Estrelas, 4150-762 Porto, Portugal\label{Porto} \and 
University of Warwick, Department of Physics, Gibbet Hill Road, Coventry, CV4 7AL, UK \label{Warwick} \and 
Institut d'Astrophysique de Paris, UMR7095 CNRS, Universit\'{e} Pierre \& Marie Curie, 98 bis boulevard Arago, 75014 Paris, France\label{IAP} \and 
Observatoire de Haute-Provence, Universit\'{e} d'Aix-Marseille \& CNRS, 04870 Saint Michel l'Observatoire, France\label{OHP} \and
Observatoire de Genève, Université de Genève, 51 Chemin des Maillettes, CH-1290 Sauverny, Switzerland\label{Geneve} \and
Astrophysics Research Centre, Queen's University Belfast, University Road, Belfast BT7 1NN, UK\label{Belfast} \and
Centre for Exoplanet Science, SUPA, School of Physics and Astronomy, University of St Andrews, St Andrews, KY16 9SS, UK\label{StAndrews} \and
INAF - Osservatorio Astrofisico di Torino, Via Osservatorio 20, 10025 Pino Torinese, Italy\label{INAF} \and
INAF - Osservatorio Astrofisico di Catania, Via S. Sofia 78, I-95123 Catania, Italy\label{INAF-Catania} \and
Cavendish Laboratory, JJ Thompson Avenue, CB3 0HE, Cambridge, UK\label{Cambridge} \and
Institute of Astronomy, Madingley Road, CB3 0HA, Cambridge, UK\label{Cambridge-IA} \and
Astrophysics Group, Keele University, Staffordshire, ST5 5BG, UK\label{Keele} \and
School of Physics \& Astronomy, University of Birmingham, Edgbaston, Birmingham, B15 2TT, UK\label{Birmingham} \and
Instituto de Astrosf\'isica de Canarias (IAC), 38205 La Laguna, Tenerife, Spain\label{IAC} \and
Departamento de Astrof\'isica, Universidad de La Laguna (ULL), 38206 La laguna, Tenerife, Spain\label{ULL} \and
Max Planck Institue for Astronomy, K\"{o}nigstuhl 17, 69117 Heidelberg, Germany\label{MaxPlanck} \and
School of Physical Sciences, The Open University, Milton Keynes, MK7 6 AA, UK\label{OpenUniv} \and
Institute for Astronomy, Astrophysics, Space Applications and Remote Sensing, National Observatory of Athens, 15236 Penteli, Greece\label{Greece} \and 
Department of Physics, University of Rome Tor Vergata, Via della Ricerca Scientifica 1, 00133 Roma, Italy\label{Rome}\\
}

\date{Received date / Accepted date }
           
\abstract{To investigate the origin of the features discovered in the exoplanet population, the knowledge of exoplanets' mass and radius with a good precision ($\lesssim 10\,\%$) is essential. To achieve this purpose the discovery of transiting exoplanets around bright stars is of prime interest. In this paper, we report the discovery of three transiting exoplanets by the SuperWASP survey and the SOPHIE spectrograph with mass and radius determined with a precision better than 15\,\%. WASP-151b and WASP-153b are two hot Saturns with masses, radii, densities and equilibrium temperatures of $0.31_{-0.03}^{+0.04}\,\mathrm{M_J}$, $1.13_{-0.03}^{+0.03}\,\mathrm{R_J}$, $0.22_{-0.02}^{+0.03}\,\rho_{\mathrm{J}}$ and $1,290_{-10}^{+20}\,\mathrm{K}$, and $0.39_{-0.02}^{+0.02}\,\mathrm{M_J}$, $1.55_{-0.08}^{+0.10}\,\mathrm{R_J}$, $0.11_{-0.02}^{+0.02}\,\rho_{\mathrm{J}}$ and $1,700_{-40}^{+40}\,\mathrm{K}$, respectively. Their host stars are early G type stars (with $\mathrm{magV} \sim 13$) and their orbital periods are 4.53 and 3.33 days, respectively. WASP-156b is a Super-Neptune orbiting a K type star ($\mathrm{magV} = 11.6$) . It has a mass of $0.128_{-0.009}^{+0.010}\,\mathrm{M_J}$, a radius of $0.51_{-0.02}^{+0.02}\,\mathrm{R_J}$, a density of $1.0_{-0.1}^{+0.1}\,\rho_{\mathrm{J}}$, an equilibrium temperature of $970_{-20}^{+30}\,\mathrm{K}$ and an orbital period of 3.83 days. 
The radius of WASP-151b appears to be only slightly inflated, while WASP-153b presents a significant radius anomaly compared to the model of \citet{2008A&A...482..315B}. 
WASP-156b, being one of the few well characterised Super-Neptunes, will help to constrain the still debated formation of Neptune size planets and the transition between gas and ice giants. The estimates of the age of these three stars confirms an already observed tendency for some stars to have gyrochronological ages significantly lower than their isochronal ages. We propose that high eccentricity migration could partially explain this behaviour for stars hosting a short period planet.
Finally, these three planets also lie close to (WASP-151b and WASP-153b) or below (WASP-156b) the upper boundary of the Neptunian desert. Their characteristics support that the ultra-violet irradiation plays an important role in this depletion of planets observed in the exoplanet population.
}


\titlerunning{Discovery of WASP-151b, WASP-153b and WASP-156b}
\authorrunning{O. Demangeon et al.}

\maketitle 

\section{Introduction}\sectlabel{intro}

The successful harvest of exoplanets \citep[see for example exoplanet.eu,][]{2011A&A...532A..79S} during the last two decades completely metamorphosed the field of exoplanet science. The initial assumption that the solar system was a typical example of planetary systems is long gone \citep[as stated by][]{2012NewAR..56...19M}. The Kepler mission \citep{2010Sci...327..977B} delivered 4,496 transiting planetary candidates, including 2,248 confirmed planets (according to the NASA Exoplanet Archive, \url{http://exoplanetarchive.ipac.caltech.edu/}, August 2017). This sample revealed various features of the exoplanet population demonstrating the necessity of a very large sample to encompass the exoplanets' diversity \citep[see][for a recent review]{2017PAPhS.161...38B}.
One of many surprising results from Kepler is that the orbital distance of exoplanets appears to be nearly random regardless of their size \citep[e.g.][]{2014ApJ...790..146F}. One striking exception to this observation is the so called sub-jovian desert or short period Neptunian desert \citep[e.g][]{2016A&A...589A..75M, 2016ApJ...820L...8M, 2014ApJ...783...54K, 2011ApJ...727L..44S}. It corresponds to a depletion of planets at short orbital periods ($P < 10\,\mathrm{days}$) with masses or radius between super-Earth and sub-jovian planets (see \fig{Neptuniandesert}). One possible explanation for this desert is the strong irradiation (bolometric and in particular extreme ultra-violet) from the parent star at those short orbital distances, especially at the early stages of the star's life. The strong stellar irradiation might have striped away the atmosphere of sub-jovian planets which had quickly migrated to the vicinity of their parent star and were not massive enough to retain their atmosphere, only leaving a super-Earth size core \citep[e.g.][]{2016NatCo...711201L, 2007A&A...461.1185L}. 
The mechanism responsible for the presence of giant planets in the vicinity of their parent star is still debated. However, the discovery by \citet{2016Natur.534..658D} of a Super-Neptune size planet orbiting close to  a 5-10\,Myr old star suggests that high eccentricity migration \citep[e.g.][]{1996Sci...274..954R,2007ApJ...669.1298F} is unlikely for this system (the tidal circularisation happening at longer timescales) and only leaves disk migration \citep[e.g.][]{1996Natur.380..606L,1997Icar..126..261W} and in-situ formation as possible scenarios.
Understanding the origin of the Neptunian desert could thus change our vision of gas and ice giant planet formation and evolution.

Unfortunately a large fraction of the planets discovered by Kepler surrounding the Neptunian desert don't have an accurate (precision $\lesssim 10\,\%$\footnote{The exact precision required is difficult to assess, but a precision of 20 to 30\,\% on the planetary density is usually required to be able to discriminate between the main families of planets \citep[see for example][]{CHEOPSRedBook, 2009ApJ...693..722G}. This correspond to an uncertainty on the radius of roughly 10\,\%.
}) determination of their mass and radius due to the faintness of their parent star. In this context ground based transit photometry surveys like SuperWASP \citep{2006PASP..118.1407P}, targeting bright stars, are essential contributors. In this paper, we present the WASP and SOPHIE discovery of two hot Saturns and one warm super Neptune, with mass and radius measured with a precision better than 15\,\%, and discuss their impact on the formation and evolution theories of ice and gas giants. In \sect{obs}, we describe the photometric and radial velocity observations acquired on the three systems. In \sect{results}, we present our analysis of the data with the resulting stellar and planetary parameters. Finally in \sect{discconclusion}, we discuss the nature and composition of these planets and their impact on planet formation and evolution theory with a focus on the migration of the hot giant planet population and the upper boundary of the Neptunian desert.

\section{Observations}\sectlabel{obs}

\subsection{Discovery: WASP}\sectlabel{dicoveryobs}
The Wide Angle Search for Planets (WASP) operates two robotic telescope arrays, each consisting of eight Canon 200m, f/1.8 lenses with e2v $2048 \times 2048$, Peltier-cooled CCDs, giving a field of view of $7.8 \times 7.8$ degrees and a pixel scale of 13.7\arcsec \citep{2006PASP..118.1407P}. SuperWASP is located at the Roque do los Muchachos Observatory on La Palma (ORM - ING, Canary Islands, Spain), while WASP-South is located at the South African Astronomical Observatory (SAAO - Sutherland, South Africa). Each array observes up to eight pointings per night with a typical cadence of $8$\,min and an exposure time of $30$\,seconds, with each pointing being followed for roughly five months per observing season.  In January 2009, SuperWASP received a significant system upgrade that improved our control of red noise sources such as temperature-dependent focus changes \citep{2011AA...525A..54B, 2011AA...531A..40F}, leading to substantially improved data quality.

All WASP data are processed by the custom-built reduction pipeline described in \citet{2006PASP..118.1407P}, producing one light curve per observing season and camera. These light curves are passed through the SysRem \citep{2005MNRAS.356.1466T} and TFA \citep{2005MNRAS.356..557K} de-trending algorithms to reduce the effect of known systematic signals, before a search for candidate transit signals is performed using a custom implementation of the Box Least-Squares algorithm (BLS; \citealt{2002AA...391..369K}), as described in \citet{2006MNRAS.373..799C, 2007MNRAS.380.1230C}. Once candidate planets have been identified, a series of multi-season, multi-camera analyses are carried out to confirm the detection and improve upon initial estimates of the candidates' physical and orbital parameters, which are derived from the WASP data in conjunction with publicly available catalogues (e.g. UCAC4, \citealt{2013AJ....145...44Z}; 2MASS, \citealt{2006AJ....131.1163S}). These additional analyses are essential for rejection of false positives, and for identification of the best candidates. This process allowed to detect three transit planets that we will now introduce. 

{\small 1SWASPJ231615.22+001824.5 (2MASS23161522+0018242)}, hereafter WASP-151, lies very close to the celestial equator and is thus visible to both WASP arrays. A total of $45,945$ data were obtained between 2008-06-12 and 2012-11-28, $16,375$ from SuperWASP and $29,570$ by WASP-South.
A search for periodic modulation in the WASP light curves, such as might be caused by stellar activity or rotation, was carried out using the method of \citet{2011PASP..123..547M}. No significant periodicity was identified, and we place an upper limit of $2$\,mmag on the amplitude of any modulation.
During these observations a total of 195 transits were covered of which 27 were full or quasi-full events.
The WASP data show a periodic reduction in stellar brightness of approximately $0.01$\,mag, with a period of roughly $4.5$\,days, a duration of approximately $3.7$\,hours, and a shape indicative of a planetary transit.
The WASP thumbnails of WASP-151 show some contamination from a background galaxy about 20" from the target and thus within our first aperture. The galaxy is about 3 magnitude fainter in V than our target. We calculated a dilution factor for WASP-151 of about 1\% and thus negligible when considering WASP data.

{\small1SWASPJ183702.97+400107.4 (2MASS18370297+4001073)}, hereafter WASP-153, is our second transiting planet host. $42,349$ photometric measurements were made by SuperWASP between 2004-05-14 and 2010-08-24, with no observations by WASP-South owing to the high declination of the target ($+40^o$). We found no significant periodic modulation, and we place an upper limit of $1.5$\,mmag on the amplitude of any such light curve variation.
There are a total of 688 transits observed of which 54 are good events\footnote{Good events refers to full transit observations which didn't suffer from obvious deformations due to the conditions of observation.
}. Our BLS searches identified the signature of a candidate transiting planet on a $3.3$\,days orbit, in the form of a periodic $0.006$\,mag, $3$\,hours reduction in stellar brightness.

{\small1SWASPJ021107.61+022504.8 (2MASS02110763+0225050)}, hereafter WASP-156, is our third and last transiting planet host. We again found no significant periodic modulation, and we place an upper limit of $1$\,mmag on the amplitude of any such light curve variation. As with WASP-151, the equatorial declination of WASP-156 allows both WASP arrays to monitor the star for flux variations. $22809$ flux measurements were made, $13481$ by SuperWASP and $9328$ by WASP-South.
A total of 230 transits were observed of which 23 are good events\footnotemark[2].
A $2.3$\,hours long, $0.007$\,mag reduction in brightness was found to repeat on a $3.8$\,days period with a typical planetary transit-like shape.

\subsection{Photometric follow-up}\sectlabel{fupphotobs}

\begin{table}[!htb]
\tiny
\caption{\tablabel{photobssummary}Summary of the photometric observation of WASP-151, WASP-153 and WASP-156.}
\begin{tabular}{llll}
\hline
 Date & Instrument & Filter & Comment \\ 
\hline
\multicolumn{4}{c}{WASP-151b}\\
\hline
06/2008$\rightarrow$11/2012	& WASP		& Johnson R	& detection \\
03/09/2015 				& IAC80		& Johnson R 	& full transit \\
01/11/2015 				& IAC80		& Johnson R 	& full transit \\
15/06/2016 				& TRAPPIST	& Sloan z 		& full transit \\
04/09/2016 				& EulerCam	& NGTS 		& partial transit \\
24/10/2016 				& EulerCam	& NGTS 		& full transit \\					    
12/2016$\rightarrow$03/2017	& K2			& Kepler		& 13 full transits \\
\hline
\multicolumn{4}{c}{WASP-153b}\\
\hline
05/2004$\rightarrow$08/2010	& WASP		& Johnson R	& detection \\
17/07/2015 				& Liverpool	& Johnson R 	& partial transit \\
05/08/2017 				& RISE-2	& V+R 	& full transit \\
\hline
\multicolumn{4}{c}{WASP-156b}\\
\hline
07/2008$\rightarrow$12/2010	& WASP		& Johnson R	& detection \\
29/12/2014 				& EulerCam	& Gunn z 		& full transit \\
07/11/2016 				& EulerCam	& Gunn r		& full transit \\
27/12/2016 				& NITES		& Johnson I 	& partial transit \\
\hline 
\end{tabular} 	
\end{table}

\subsubsection{Ground-based photometric follow-up observations}\sectlabel{gbfupphotobs}
The WASP consortium has access to multiple observing facilities that can be used to obtain additional in-transit photometric observations. These follow-up light curves are used to confirm the presence of the candidate signal, particularly useful in the case of unreliable initial ephemerides, and are also used to improve the accuracy of our light curve modelling, and to constrain the system parameters more precisely. A list of the follow-up photometric observations for our three planets is presented in Table\,\ref{tab:photobssummary}. 

\paragraph{WASP-151b}
Two full transits of WASP-151 were observed on 2015-09-03 and 2015-11-01 with the CAMELOT camera of the $0.82$\,m ($f/11.3$) IAC-80 telescope, which is operated on the island of Tenerife by the Instituto de Astrof{\'i}sica de Canarias (IAC) at the Spanish Observatorio del Teide. CAMELOT has a $2048 \times 2048$\,pixel CCD with a scale of $0.304$\arcsec\,pixel$^{-1}$ and a $10.6$\,\arcmin field-of-view. Images were bias and flat-field corrected using standard techniques.

An additional full transit was observed on 2016-06-15 with the robotic $0.6$\,m TRAnsiting Planets and PlanetesImals Small Telescope (TRAPPIST; \citealt{2011Msngr.145....2J,2011EPJWC..1106002G}) at the La Silla Observatory operated by the European Southern Observatory (ESO). TRAPPIST is equipped with a thermoelectrically-cooled 2K\,$\times$\,2K CCD with a pixel scale of $0.65$\arcsec, giving a $22\arcmin\times22\arcmin$ field of view. A Sloan-z$^\prime$ filter was used for the transit observations of this system, during which the positions of the stars on the chip were maintained to within a few pixels thanks to a software guiding system that regularly derives an astrometric solution for the most recently acquired image and sends pointing corrections to the mount if needed. After carrying out bias, dark, and flat-field corrections we extract stellar fluxes from our images using the IRAF\footnote{IRAF is distributed by the National Optical Astronomy Observatories, which are operated by the Association of Universities for Research in Astronomy, Inc., under cooperative agreement with the National ScienceFoundation.}/ DAOPHOT aperture photometry software \citep{1987PASP...99..191S}. Several sets of reduction parameters were tested on stars of similar brightness to WASP-151, from which we selected the set giving the most precise photometry. After a careful selection of reference stars, the transit light curves were finally obtained using differential photometry. 

The $1.2$\,m Swiss telescope using EulerCam \citep{2012AA...544A..72L}, also at La Silla, observed a full transit of WASP-151b on 2016-10-24 and a partial transit on 2016-10-24. In both cases, a filter with a central wavelength of $698$\,nm and an effective bandwidth of $312$\,nm was used; this filter is the same as that used by the Next Generation Transit Survey (NGTS; \citealt{2013EPJWC..4713002W, 2014IAUS..299..311W}). The Swiss telescope employs an absolute tracking system which matches point sources in each image with a catalogue and adjusts the telescope's pointing between exposure to compensate for drift. In this manner, the pixel position of the star is maintained throughout. All data were reduced as outlined in \citet{2012AA...544A..72L}, and light curves were produced through differential aperture photometry. To minimise scatter in the light curves, we carefully selected the most stable field stars to use as references.
 
\paragraph{WASP-153b}
A partial transit of WASP-153b was observed on 2015-07-17 in the Johnson-R filter using the RISE instrument mounted on the robotic Liverpool Telescope (LT; \citealt{2004SPIE.5489..679S}) at ORM. RISE is equipped with a back-illuminated, frame-transfer, $1024\times1024$\,pixel CCD. Images were automatically bias, dark, and flat-field corrected by the standard RISE reduction pipeline, which uses standard IRAF routines.

A full transit was later obtained with RISE-2 mounted on the 2.3 m telescope situated at Helmos observatory in Greece on 2017 August 5. The CCD size is 1K\,$\times$\,1K with a pixel scale of $0.51$\,\arcsec\ and a field of view of 9\arcmin\,$\times$\,9\arcmin\ \citep{2010ASPC..424..426B}. The exposure time was $12\,s$ and the V$+$R filter was used. As for the previous transit observation, the images were processed standard RISE reduction pipeline.

\paragraph{WASP-156b}
A partial transit of WASP-156b was observed in the Johnson-I filter on 2016-12-27 using the Near Infra-red Transiting ExoplanetS (NITES) Telescope \citep{2014MNRAS.438.3383M}, located at ORM. NITES is a semi-robotic, $0.4$\,m ($f/10$) Meade LX200GPS Schmidt-Cassegrain telescope, mounted with a Finger Lakes Instrumentation Proline 4710 camera and a $1024\times1024$ pixel deep-depleted CCD made by e2v. The telescope has a field of view of $11\times11\arcmin$ squared, and a pixel scale of $0.66$\arcsec\,pixel$^{-1}$. Autoguiding is performed using the DONUTS algorithm \citep{2013PASP..125..548M}. After performing bias and flat-field corrections using PyRAF\footnote{PyRAF is a product of the Space Telescope Science Institute, which is operated by AURA for NASA.} and the standard routines in IRAF, aperture photometry was performed using DAOPHOT and multiple comparison stars, selected to minimise the RMS scatter in the out-of-transit light curve.

In addition to the NITES observations, EulerCam was used to observe two full transits of WASP-156b, on 2014-12-29 using a Gunn-z filter and on 2016-11-07 using a Gunn-r filter. The 2014 observations, however, are unreliable owing to large PSF variations, and stellar counts in the non-linear regime of the EulerCam CCD.

\subsubsection{K2 observations of WASP-151}\sectlabel{K2obs}

In addition to the ground-based photometric observations described in the previous sections, WASP-151 
(alias EPIC\,246441449) was observed by NASA's Kepler Space Telescope in its two-reaction wheel 
mission K2 \citep{2014PASP..126..398H} during Campaign 12. 
The observations span over $\sim79$\,days (from 15th December 2016 to 4th March 2017) apart from 5\,days 
(from 1st February 2017 to 6th February 2017) when the spacecraft was in safe mode.

Since Campaign 9, the K2 consortium releases the raw cadence data shortly after downlink from the Kepler satellite.
These data are raw, as opposed to the science cadence data like the target pixel files (\textsc{tpf}), for two main reasons\footnote{For more details on the Kepler raw and science cadence data, we refer the reader to the technical note entitled Format Information for Cadence Pixel Files available at https://archive.stsci.edu/k2/manuals/KADN-26315.pdf}.
First their format, the raw cadence data are provided as one file per cadence delivering the pixel counts for the whole focal plane as a table. In order to construct the image time series of a target, we need the pixel mapping reference file which specifies the (column, row) CCD coordinates for each value in the raw cadence data tables. 
Second, the raw cadence data are not calibrated. It means that they are not reduced with the Kepler pipeline \citep{2010SPIE.7740E..1XQ} and thus not corrected for background, dark, smearing trails, undershoot or non-linearity of the pixels response. The formatting and calibration of the raw cadence data for all the targets of a K2 campaign is a very lengthy procedure and even if the raw cadence data for Campaign 12 have been released several months ago, the calibrated \textsc{tpf} are, at this moment, still unavailable.
Therefore, to be able to benefit from the high quality light-curves of the WASP-151 system provided by the K2 mission, we decided to format and reduce ourselves the raw cadence data.

To obtain an image time series, we used the \texttt{Kadenza}\footnote{The \texttt{Kadenza} software is available on GitHub at https://github.com/KeplerGO/kadenza or on Zenodo at https://doi.org/10.5281/zenodo.344973
} software \citep{2017misc10.5281} provided by the NASA's Kepler/K2 Guest Observer Office.
Then, to extract the light-curve, we used the \texttt{Polar} software \citep{2016A&A...594A.100B} which performs a partial calibration by subtracting the background and dark values thanks to estimates obtained on the images themselves.
In parallel to the \texttt{Polar} reduction, we also reduced the image time series with the Python package \texttt{Everest} \citep{2016AJ....152..100L} to check the scientific validity of our reduction. \texttt{Everest} has been recently used to extract the light-curve of the TRAPPIST-1 system observed by K2 during the same campaign \citep{2017arXiv170304166L} and thus in the same conditions. The two light-curves are almost identical and compatible at 1 sigma giving us confidence in the scientific quality of our data reduction. 

The light-curve clearly display transit features at the ephemeris inferred from the WASP data with no sign of out-of-transit variations. A search for periodic modulation caused by stellar activity showed a tentative detection with an amplitude of 1\,{ppt} ($\sim 1\,\mathrm{mmag}$) at a period of 35 days.
We then searched the light-curve for additional transit features (apart from WASP-151b's transit). We investigated a tentative mono transit-shaped feature which proved to be an artifact due to the position-flux decorrelation technique used by \texttt{Polar}. For this decorrelation, we cut the K2 image time-series in several parts where the behavior of the pointing jitter of the Kepler satellite can be safely assumed to be 1 dimensional \citep[for more details see][]{2016A&A...594A.100B}. The mono transit-shaped feature was appearing precisely at the junction of two of those parts. A slight change of the location of the cut made the feature disappear. Finally, no additional transit features was detected. 

For the analysis, we only kept intervals of 2 times the transit duration before and after each transit of WASP-151b. The phase-folded \texttt{Polar}-K2 light-curve of WASP-151 is shown in the bottom panel of \fig{W151LCdatacomp}. 

\subsection{Spectroscopic follow-up}\sectlabel{fupspectroobs}

The spectroscopic follow-up of these three candidates was mainly performed with 
SOPHIE, the spectrograph dedicated to high-precision radial velocity 
measurements at the 1.93-m telescope of the Haute-Provence Observatory, 
France \citep{2009A&A...505..853S}. For two systems, it was also 
complemented by radial velocities obtained with the CORALIE spectrograph 
at the 1.2-m Euler-Swiss telescope at La Silla \citep{2000A&A...354..99S}, Chile.
The first goal of these spectroscopic observations is to establish the planetary nature of the transiting candidates found in photometry (see \sect{validation})
The second goal is to characterize the secured planets by measuring in particular their masses and orbital eccentricities (see \sect{trrvanalysis}).

\subsubsection{Description of the observations}

SOPHIE was used in High-Efficiency mode with a resolving power 
$R=40\,000$ to increase the throughput for these faint stars.
The exposure times ranged from 400 to 2200~sec depending on the targets, and they 
were adjusted as a function of the weather conditions to keep the 
signal-to-noise ratio as constant as possible for any given star.
The spectra were extracted using the SOPHIE pipeline, and the radial velocities 
were measured from the weighted cross-correlation with numerical masks
characteristic of the spectral type of the observed star \citep{1996A&AS...119..373S, 2002A&A...338..632S}. 
We adjusted the number of spectral orders used in the cross-correlation 
to reduce the dispersion of the measurements. Some spectral domains  
are noisy (especially in the blue part of the spectra) and using them would have degraded the 
accuracy of the radial-velocity~measurement.

The error bars on the radial velocities were computed from the cross-correlation 
function using the method presented by \citet{2010A&A...523..A88}. 
Some spectra were contaminated by moonlight. 
Following the method described in \citet{2008MNRAS.386.1576D} and
\citet{2009A&A...481..52S}, we estimated and corrected for the moonlight contamination 
by using the second SOPHIE fiber aperture, which is targeted on the sky, while the first 
aperture points toward the star. This results in radial velocity corrections up to 40~m/s, 
and below 40~m/s in most of the cases.
Removing these points does not significantly modify the orbital~solutions.

The CORALIE spectrograph has a resolution of $\sim 60,000$. The observing strategy is made to ensure that observations are taken exclusively without Moon contamination and the second fiber is used to obtain a simultaneous calibration. Prior to April 2015 the calibration was done with a Thorium-Argon lamp, but since then it is done with a Fabry-P\'erot unit. The reduction of the spectra and the production of the radial velocities proceed in a fashion very similar to the procedure applied to SOPHIE data.

The radial velocity measurements are reported in Tables~\tabalt{rvobsWASP151}, \tabalt{rvobsWASP153}, \tabalt{rvobsWASP156} and are displayed in Figures~\figref{W151RVdatacomp}, \figref{W153RVdatacomp} and \figref{W156RVdatacomp}
together with their Keplerian fits and the residuals.

\subsubsection{Validation of the planetary nature}
\sectlabel{validation}

The transit photometry method suffers from a high rate of false positives. Eclipsing binaries (\textsc{eb}), background eclipsing systems (\textsc{bes}) and hierarchical triple systems (\textsc{hts}) can mimic the transit of a planet orbiting the target star and induce an erroneous identification of the nature and parameters of the transiting system \citep[e.g.][]{2014MNRAS.441..983D, 2011ApJ...727...24T}.
Whenever it is possible, radial velocity measurements are used to rule out these false positive scenarios and validate the planetary nature of the transiting object.
This validation is made in several steps.
\begin{enumerate}
\item The inspection of the spectra allows to identify double lines spectrum which are sign of spectroscopic binaries (SB2) or \textsc{bes}/\textsc{hts} where the contaminating eclipsing system has a similar brightness than the target star.
\item Phase-folding the data at the period inferred from the transits allows to estimate the amplitude of the \textsc{rv} signal at this period. Assuming that this amplitude is due to the reflex motion of the target star, it allows to estimate the mass of the gravitationally bound companion and to identify single line binaries (SB1).
\end{enumerate}
If those two steps are successfully passed, the \textsc{eb} scenario can be ruled-out\footnote{The following steps rely on the fact that a significant \textsc{rv} variation is detected during the second step. If it's not the case, the only remaining solution is often to assess the nature of the transiting signal through probabilistic validation. Few softwares exist to perform this probabilistic validation: \textsc{blender} \citep{2011ApJ...727...24T}, \textsc{pastis} \citep{2014MNRAS.441..983D} and under more restrictive assumptions \textsc{vespa} \citep{2015ascl.soft03011M, 2012ApJ...761....6M}.
}.
For our three planetary candidates, none of the measurements show double lines. Furthermore, they show variations in phase with the SuperWASP transit ephemeris and with semi-amplitudes between 20 and 40~m/s, implying companion masses below 0.4~Jupiter mass. Therefore, we can exclude the \textsc{eb} hypothesis for our three cases. The remaining false positive scenarios are thus \textsc{bes} and \textsc{hts} with faint\footnote{As described in the first step, we can also exclude \textsc{bes} and \textsc{hts} configurations involving bright contaminating eclipsing systems up to a flux ratio between the contaminant and the greater than  $\sim 1\,\%$.
} contaminating eclipsing systems.  
\begin{enumerate}
\setcounter{enumi}{2}
\item Extracting the radial velocities using masks corresponding to different spectral types allows to identify some cases of \textsc{hts}/\textsc{bes} where the star responsible for the \textsc{rv} signal has a different spectral type than the target star. In such a case, the \textsc{rv} amplitude will vary significantly with the mask used \citep[e.g.][]{2002A&A...392..215S}.
\item If the \textsc{rv} signal observed is due to a \textsc{hts}/\textsc{bes}, it will display variation in the cross-correlation function bisector span (\textsc{bs}) correlated with the \textsc{rv} signal \citep{2015MNRAS.451.2337S}. It is thus important to properly assess the correlation between \textsc{rv} and \textsc{bs}, since a significant correlation would exclude the planetary hypothesis\footnote{A correlation can be explained by a \textsc{bes} or a \textsc{hts} but also by stellar activity \citep[e.g.][]{2001A&A...379..279Q}.
}.
\end{enumerate}
For our three planetary candidates, radial velocities were measured using different stellar masks (F0, G2, and K5) and produced variations with similar amplitudes. Furthermore, \fig{bisrv} shows the correlation diagram of the \textsc{rv} and \textsc{bs} signal along with the \textit{posterior} probability density function of the correlation coefficient, obtained with the method and tools described in \citet{2016OLEB...46..385F}. The values and $95\,\%$ confidence intervals that we obtained are $0.19_{-0.31}^{+0.28}$, $-0.01_{-0.26}^{+0.26}$ and $-0.13_{-0.24}^{+0.25}$ for WASP-151b, WASP-153b and WASP-156b respectively, meaning that no significant correlation is detected.

The final step that is rarely performed, when a \textsc{rv} variation is significantly detected, is to check whether or not a correlation could have been detected assuming that the \textsc{rv} variation is due to a \textsc{hts} or a \textsc{bes}.
\citet{2015MNRAS.451.2337S} described in details the expected \textsc{rv} and \textsc{bs} signals for \textsc{hts}/\textsc{bes}. The exact degree of correlation and the exact amplitude ratio of \textsc{bs} over \textsc{rv} depend on the following factors: flux ratio, full width at half maximum  of the cross-correlation functions (\textsc{fwhm}), mean radial velocity difference ($\phi$) and spectral types.
However in most configurations\footnote{According to \citet{2015MNRAS.451.2337S}, the only \textsc{hts}/\textsc{bes} configuration which might produce a \textsc{rv} signal with a comparatively low \textsc{bs} signal is when the \textsc{fwhm} or the target and the contaminating systems are similar and $\phi$ is low compared to this \textsc{fwhm} value. Given the \textsc{fwhm} of $\sim 5\,\kms$ of our observation, this is only possible a specific kind of \textsc{hts} system.
}, to be able to produce the $\sim 30\,\ms$ \textsc{rv} variation that we observe, the associated \textsc{bs} signal must have an amplitude equal to a significant fraction of the \textsc{rv} signal. This in turn implies that the ratio of the dispersion over the average error bar of the \textsc{bs} measurements ($\frac{\mathrm{std(\textsc{bs})}}{\langle \sigma_{\textsc{bs}} \rangle}$) has to be greater than 1.
Consequently, we computed $\frac{\mathrm{std(\textsc{bs})}}{\langle \sigma_{\textsc{bs}} \rangle}$ for our three stars and this ratio is compatible with one in all cases (see first column of \tab{BSdisp}). This implies that the dispersion of the \textsc{bs} values can be explained by the measurement uncertainties solely and discards cases where the additional \textsc{bs} signal due to the \textsc{hts}/\textsc{bes} could have been detected. To quantify these cases, we computed the maximum fraction of the \textsc{rv} amplitude that the \textsc{bs} signal can have without producing a 2 sigma departure from 1 of $\frac{\mathrm{std(\textsc{bs})}}{\langle \sigma_{\textsc{bs}} \rangle}$ (see second column of \tab{BSdisp}).

With \tab{BSdisp}, we can identify the configurations of \textsc{hts}/\textsc{bes} that are excluded by our correlation and \textsc{bs} dispersion analyses, given the number and the precision of our \textsc{rv} and \textsc{bs} measurements. We thus conclude that for our three stars, we would have been able to detect the increase in the dispersion of the \textsc{bs}, and thus the correlation between \textsc{rv} and \textsc{bs} associated with the presence of most \textsc{hts}/\textsc{bes} configurations. We are thus confident that the most likely explanation for our transits and \textsc{rv} signals is a planet orbiting the target stars.

\begin{table}[htb]
\caption{\tablabel{BSdisp}Analysis of the dispersion of the bisector span.}
\begin{tabular}{lcc}
\hline
 Star & $\frac{\mathrm{std(\textsc{bs})}}{\langle \sigma_{\textsc{bs}} \rangle}$ & max($\frac{\textsc{bs}}{\textsc{rv}}$) [$\%$] \\[3pt] 
\hline
\hline \\[-4pt]
WASP-151 	& $0.93 \pm 0.15$	& $84$ \\[3pt]
WASP-153 	& $1.15 \pm 0.12$	& $22$ \\[3pt]
WASP-156 	& $1.03 \pm 0.11$	& $48$ \\[3pt]
\hline
\end{tabular}
\\
\textbf{Note}: std(\textsc{bs}) indicates the standard deviation of the \textsc{bs} measurements. $\langle \sigma_{\textsc{bs}} \rangle$ indicates the average error bar on the individual \textsc{bs} measurements. Max($\frac{\textsc{bs}}{\textsc{rv}}$) is the maximum fraction of the \textsc{rv} amplitude observed that the \textsc{bs} signal can have without producing a value of $\frac{\mathrm{std(\textsc{bs})}}{\langle \sigma_{\textsc{bs}} \rangle}$ which is significantly superior to 1 (see \sect{validation} for more details).
\end{table}

\begin{figure*}[t!] 
 \centering
    \begin{minipage}{0.33\textwidth}
        \centering
        \includegraphics[width=\textwidth]{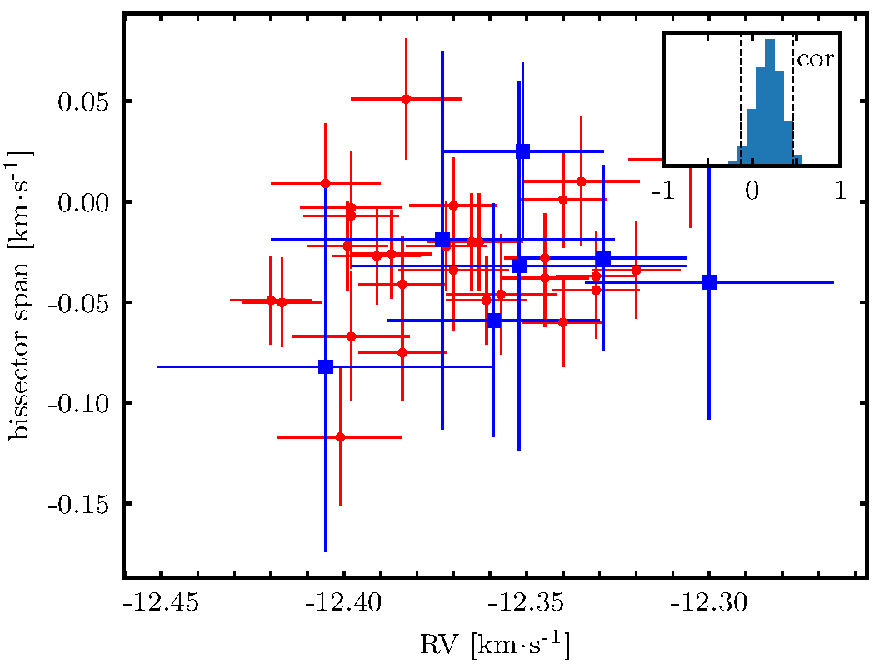}
    \end{minipage}\hfill
    \begin{minipage}{0.33\textwidth}
        \centering
        \includegraphics[width=\textwidth]{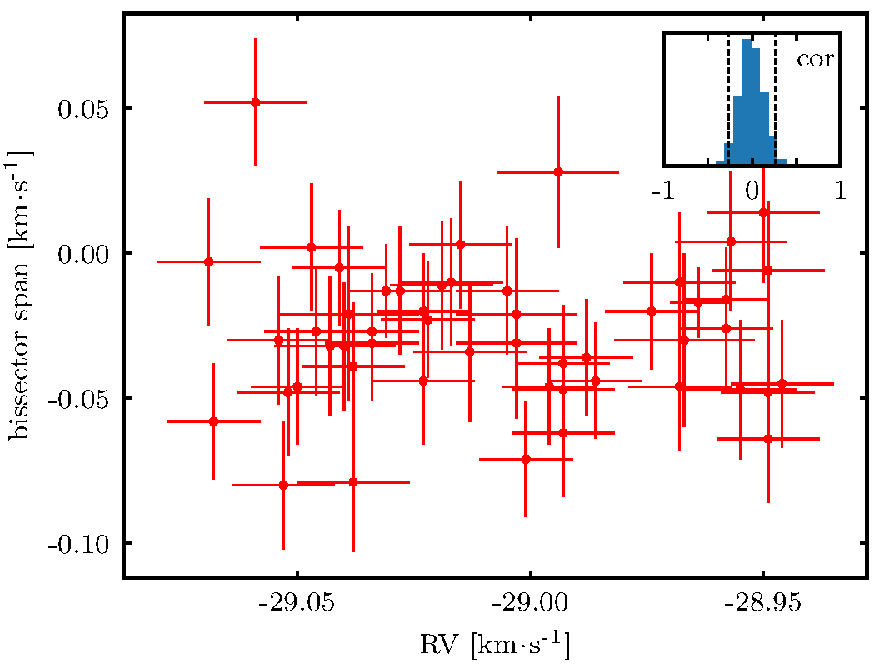}
    \end{minipage}
    \begin{minipage}{0.33\textwidth}
        \centering
        \includegraphics[width=\textwidth]{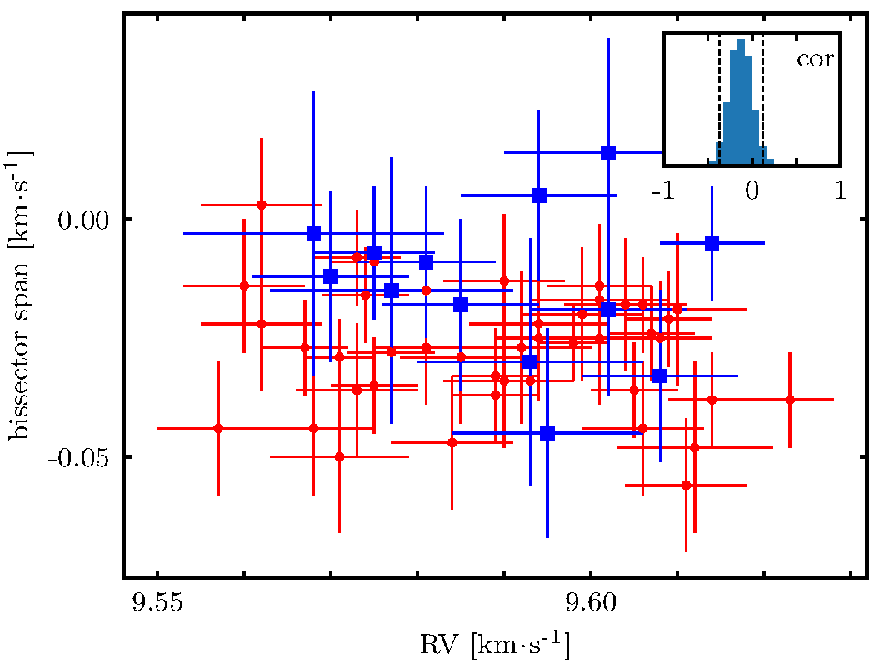}
    \end{minipage}
	\caption{\figlabel{bisrv}Bisector span as a function of the radial velocities with 1-$\sigma$\,error bars for WASP-151, 153 and 156 (from left to right). 
SOPHIE data are the red circles; CORALIE data are the blue squares.
The ranges here have the same extents in the $x$- and $y$-axes. For each star, the posterior probability function of the correlation coefficient is displayed in an insert located in the upper left corner.}
\label{fig_bisectors}
\end{figure*}

\section{Results}\sectlabel{results}

\subsection{Stellar Parameters from spectroscopy}\sectlabel{stelparres}

A total of 26, 46, and 40 individual SOPHIE spectra of WASP-151, WASP-153 and WASP-156 were co-added to produce a single spectrum with a typical S/N of around 50:1, 50:1 and 70:1, respectively. We used here only the spectra without moonlight contamination; it enabled a sufficiently high signal-to-noise ratio to be reached with $\mathrm{R}=40000$, and prevented any possible contamination in the spectra.

The standard pipeline reduction products were used in the analysis, which was performed using the methods given in \citet{2013MNRAS.428.3164D}. The effective temperature (\teff) was determined from the excitation balance of the Fe~{\sc i} lines. 
The ionisation balance of Fe~{\sc i} and Fe~{\sc ii} was used as the surface gravity (\logg) diagnostic. 
The metallicity (\feh) was determined from equivalent width measurements of several unblended lines. They are more accurate and agree with the measurements secured from the cross-correlation function following \citet{2010A&A...523..A88}.
A value for microturbulence (\mictrb) was determined from Fe lines by requiring that there is no slope between the abundance and the equivalent width.
The error estimates for \mictrb\ include the uncertainties in \teff\ and \logg, as well as the scatter due to the measurement and the atomic data uncertainties.
Values for macroturbulence (\mactrb) were determined from the calibration of \citet{Doyle14}, however the value for WASP-156 is extrapolated from the calibration as this star is not within the correct temperature range. With the \mactrb\ fixed to the calibration value, the projected stellar rotation velocity (\vsini) was determined by fitting the profiles of several unblended lines. Here again, the \vsini\ values agree with those obtained from the cross-correlation function following \citet{2010A&A...523..A88}.

Lithium is detected in WASP-151 and WASP-153, with an equivalent width of 17 m\AA\ and 98 m\AA, 
corresponding to an abundance  $\log A$(Li)  of 1.73 $\pm$ 0.05 and  2.77 $\pm$ 0.05 respectively. 
This implies an age of several Gyr and several Myr respectively. 
There is no significant detection of lithium in WASP-156, with an equivalent
width upper limit of 11 m\AA, corresponding to an abundance upper limit of $\log
A$(Li) $<$ 0.2. This implies an age of at least 500 Myr \citep{2005A&A...442..615S}.

The rotation rate ($P = 14.8 \pm 4$~d) implied by the {\vsini} gives a
gyrochronological age of $\sim1.80^{+2.03}_{-1.00}$~Gyr using the
\cite{2007ApJ...669.1167B} relation for WASP-151.  Similarly, the rotation rate of 
$P = 11.7 \pm 2$~d gives an age of $\sim 1.21^{+1.19}_{-0.60}$ Gyr for WASP-153, 
and the rotation rate of $P = 12.6 \pm 4$~d gives an age of $\sim 0.58^{+0.51}_{-0.31}$ Gyr for WASP-156.

Finally from \teff, \logg\ and \feh, we inferred stellar mass and radius estimates using the \cite{2010A&ARv..18...67T} calibration.
The parameters and error bars obtained from this analysis are listed in the section stellar parameters of \tab{syspar}.

\subsection{System Parameters}\sectlabel{sysparres}

\subsubsection{Transit and RV analysis}\sectlabel{trrvanalysis}

We followed the same method to perform the parameters' inference for the three systems. We analysed jointly all the radial velocity and photometric datasets available for a given system. To model the radial velocity and photometric data, we used the Python packages \texttt{ajplanet}\footnotemark[11] \citep{2016ApJ...830...43E}  and \texttt{batman}\footnote{\label{fnt:pythonpackages}Several of the Python packages used for this work are publicly available on Github: \texttt{ajplanet} at https://github.com/andres-jordan/ajplanet, \texttt{batman} at https://github.com/lkreidberg/batman, \texttt{emcee} at https://github.com/dfm/emcee, \texttt{ldtk} at https://github.com/hpparvi/ldtk}
\citep{2015PASP..127.1161K} respectively. In order to decrease the correlation between the parameters of our model and ease the fit, we adopted the parametrisation suggested by \citet{2013PASP..125...83E} with
${R_p / R_*}$ the ratio of the planet’s radius to that of the star, $P$ the orbital period, $t_c$ the planet's time of inferior conjunction, $\sqrt{e}\cos \omega_*$ and $\sqrt{e}\sin \omega_*$ where $e$ is the orbital eccentricity and $\omega_*$ is the stellar orbital argument of periastron, $K$ the radial velocity semi-amplitude, $i$ the orbital inclination, $a / R_*$ the ratio of the planet's orbital semi-major axis over the stellar radius, $v0$ the systemic radial velocity, $u$ and $v$ the two coefficients of the limb-darkening quadratic law. To this set of parameters we added a logarithmic multiplicative jitter factor ($\ln f_{\sigma}$) for each instrument to account for a possible bias in the data's error bars due to overestimated, underestimated or even non-considered sources of noise \citep[see][]{2009MNRAS.393..969B}. Finally, we added a parameter for the shift of the radial velocity zero point between two instruments ($\Delta\textrm{RV}$) and three coefficients to model a linear or quadratic variation of the out-of-transit relative flux ($\Delta F_\textsc{oot}$, $\Delta F^\prime_\textsc{oot}$ and $\Delta F^{\prime\prime}_\textsc{oot}$) when it was necessary. The final list of main parameters is $R_p / R_*$, $P$, $t_c$, $\sqrt{e}\cos \omega_*$, $\sqrt{e}\sin \omega_*$, $K$, $\cos i$, $a / R_*$, $v0$, $u$ and $v$, $\ln f_{\sigma}$, $\Delta\textrm{RV}$, $\Delta F_\textsc{oot}$, $\Delta F^\prime_\textsc{oot}$ and $\Delta F^{\prime\prime}_\textsc{oot}$.

To infer accurate values for these parameters, we used the maximum \textit{a posteriori} (\textsc{map}) estimator of the Bayesian inference framework \citep[e.g.][]{2005blda.book.....G}.
The \prior\ probability density functions (\pdf) assumed for the parameters are given by Table \tabref{priorandposteriors}. Along with the \post\ \pdf\ provided in \tab{syspar}, it allows for a qualitative assessment of the impact of the \prior\ on the inferred values.

The \prior\ on the limb darkening coefficients deserves a specific consideration. We used Gaussian \pdf s whose first two moments were defined using the Python package \texttt{ldtk}\footnotemark[11] \citep{2015MNRAS.453.3821P}. Using a library of synthetic stellar spectra, it computes the limb darkening profile of a star, observed in a given spectral bandpass (specified by its transmission curve), and defined by its \teff, \logg\ and \feh. Provided the values and error bars for these stellar parameters and the spectral bandpass, \texttt{ldtk} uses a Markov Chain Monte Carlo (\textsc{mcmc}) algorithm to infer the mean and standard deviation of the Gaussian \pdf s for the coefficients of a given limb-darkening law (quadratic in our case). \texttt{ldtk} relies on the library of synthetic stellar spectra generated by \citet{2013A&A...553A...6H}. It covers a wavelength range, from $500\,\mbox{\AA}$ to 5.5\,\textmu m , and a stellar parameter space delimited by: $2\,300\,\mathrm{K} \leq \teff \leq 12\,000\,\mathrm{K}$, $0.0 \leq \logg \leq +6.0$, $-4.0 \leq \feh \leq +1.0$, and $-0.2 \leq \afe \leq +1.2$. This parameter space is well within the requirements of this study (see Tables \tabref{photobssummary} and \tabref{syspar})

The likelihood functions used are multi-dimensional Gaussians, including logarithmic multiplicative jitter factors as described by \citet{2009MNRAS.393..969B}.
To estimate the \textsc{map} and infer reliable error bars, we explored the parameter space using an affine-invariant ensemble sampler for \textsc{mcmc} thanks to the Python package \texttt{emcee}\footnotemark[11] \citep[see][]{2013PASP..125..306F, 2012ApJ...745..198H}.
We adapted the number of walkers to the number of free parameters in our model. As a compromise between the speed and the efficiency of the exploration, we chose to use $\lceil \textrm{n}_{\textrm{free}} \times 2.5  \times 2\rceil / 2$  walkers, where $\textrm{n}_{\textrm{free}}$ is the number of free parameters and $\lceil~\rceil$ the ceiling function. This allows to have an even number of walkers which is at least 2 times ($\sim 2.5$ times) the number of free parameters, as suggested by \citet[][]{2013PASP..125..306F}.

The introduction in the model of a multiplicative jitter factor complicated the exploration of the parameter space since it introduced local maxima. For the affine-invariant ensemble sampler \textsc{mcmc} algorithm implemented by \texttt{emcee}, when different chains converge towards different disconnected maxima, the exploration becomes less efficient (the acceptance fraction of the chain decreases). Consequently, we separated the exploration into two phases. In a first exploration, we used values randomly generated from the \textit{priors} as initial values for the free parameters. This first exploration allowed us to locate several (usually two) local maxima, to extract the global maximum (the one with the highest \textit{posterior} probability) and to estimate its location and 68\,\% confidence level interval. Then we ran a second exploration to precisely sample the global maximum. For this one, the initial values were randomly generated with normal distributions whose mean and standard deviation were set accordingly to the location and width of the global maximum found by the previous step. The final best-fit values for each parameter were estimated from this second exploration after removing any residual burn-in phase with the Geweke algorithm \citep[see][]{1992clouk.book....G}. The \textsc{map} value for each parameter was finally estimated with the $50^{\textrm{th}}$ percentile of the associated marginal \post\ distribution. The extrema of the 68\,\% confidence level intervals were estimated with the $16^{\textrm{th}}$ and $84^{\textrm{th}}$ percentiles. These values are reported in \tab{syspar}.

In \tab{syspar}, we also reported the \textsc{map} and the 68\,\% confidence level interval for the secondary parameters. As opposed to the main (or jumping) parameters described in the first paragraph of this section, secondary parameters are not used in the parametrisation chosen for our modeling and are not necessary to perform the \textsc{mcmc} exploration. However, they provide quantities that can be computed from main parameter's values and are of interest to describe the system. The secondary parameters that we computed were: $\Delta F/ F$ the transit depth, $i$ the orbital inclination, $e$ the eccentricity, $\omega$ the argument of periastron, $b$ the impact parameter, $D14$ the outer transit duration (duration between the $1^{\textrm{st}}$ and $4^{\textrm{th}}$ contact), $D23$ the inner transit duration (duration between the $2^{\textrm{nd}}$ and $3^{\textrm{rd}}$ contact), $R_p$ the planetary radius, $M_p$ the planetary mass, $a$ the semi-major axis, $\tau_{\textrm{circ}}$ the timescale for the circularisation of the orbit, $F_{i}$ the incident flux on the top of the planetary atmosphere, $T_{\textrm{eq}}$ the equilibrium temperature of the planet (assuming an albedo of 0), $H$ the scale height of the atmosphere assuming a mean molecular weight of $2.2$ g/mol, $\rho_*$ the stellar mean density and \logg\ the stellar log gravity. Both $\rho_*$ and \logg\ are, in this case inferred from the transit profile\footnote{To obtain \logg, we also used the estimate of the stellar mass obtained in the next section (\sectref{stelmod}).
}. These estimates are marked with (\textit{tr.}) in \tab{syspar}.
After the full \textsc{mcmc} analysis, we computed the value of all these secondary parameters from the main parameters values and at each step of each walker of the second \texttt{emcee} exploration. Then we estimated their \textsc{map} and 68\,\% confidence level interval with the same method than the main parameters.

The specificities for the analysis of each system were:

\begin{figure}[!htb]
    \resizebox{\hsize}{!}{\includegraphics{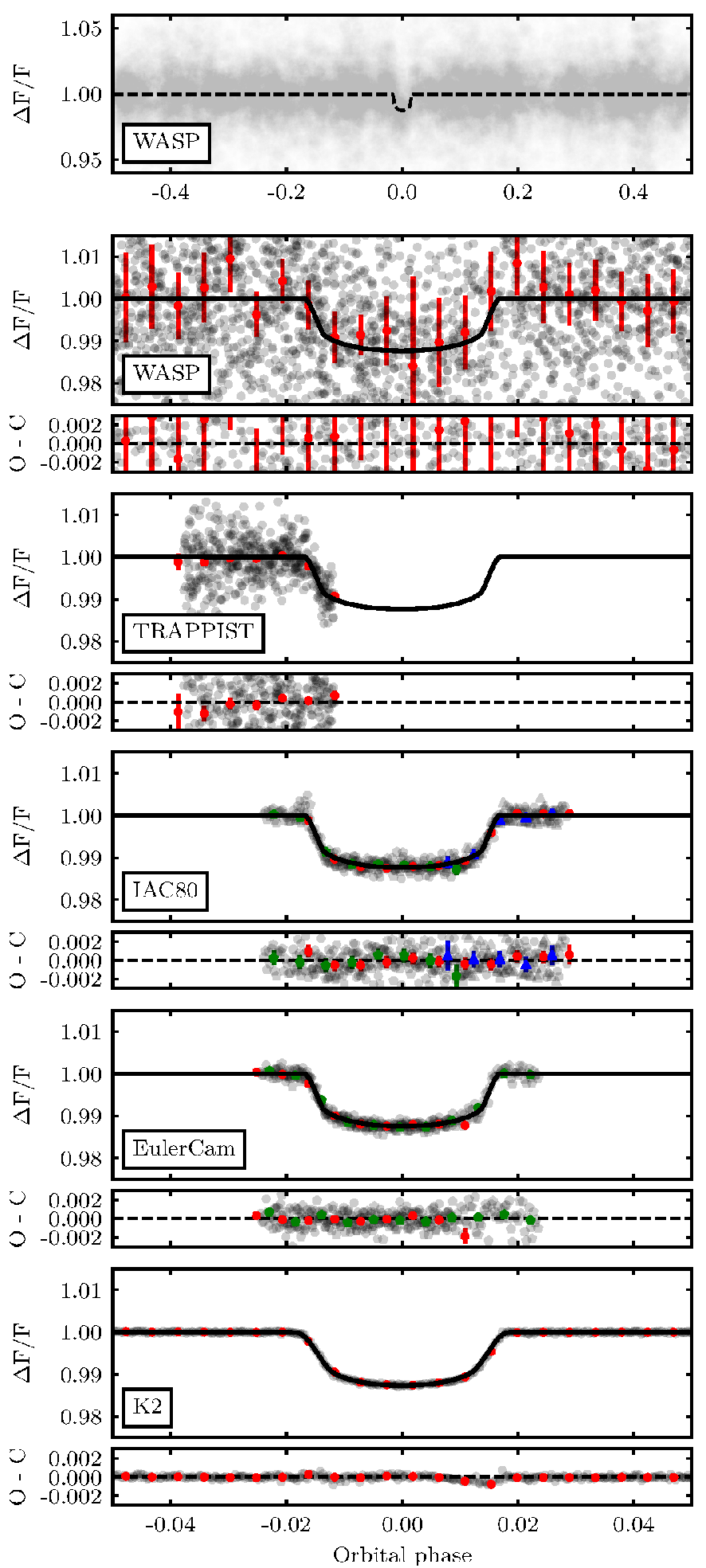}}
    \caption{\figlabel{W151LCdatacomp}Photometry of WASP-151.
The black/grey points are the data points at the original cadence of the observations, displayed without error bars for clarity. The red points corresponds to the same data points binned in phase with a bin width equivalent to 29.424 minutes (Kepler long cadence). These points are represented with their associated $1\,\sigma$ error bars. The black dashed and solid lines correspond to the best-fit model at the original and binned cadence respectively. When several datasets have been gathered with the same instrument, they are displayed on the same figure but with different symbols and colors. September EulerCam and IAC80 data are red dots, October EulerCam and first part of the November IAC80 data are green pentagons, and second part of the November IAC80 data are blue triangles.}
\end{figure}

\begin{figure}[!htb]
    \resizebox{\hsize}{!}{\includegraphics{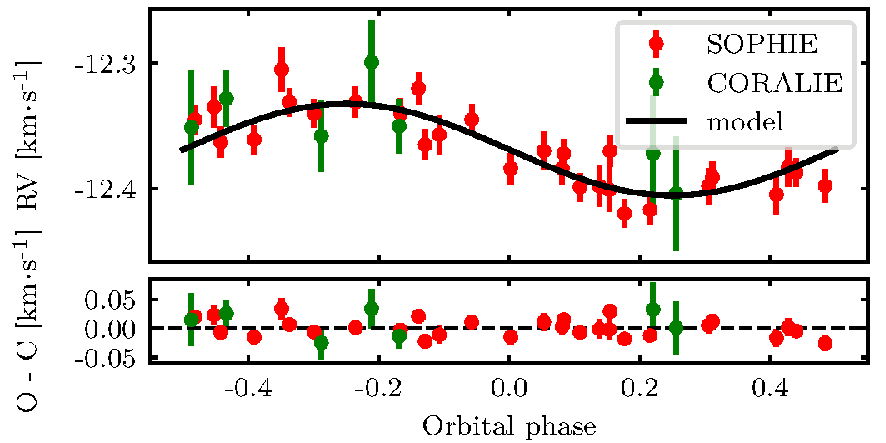}}
    \caption{\figlabel{W151RVdatacomp}Radial velocities of WASP-151. The data points are represented with their associated $1\,\sigma$ error bars.}
\end{figure}

\paragraph{WASP-151:}~The exposure times of the WASP, IAC80, EulerCam and TRAPPIST data are all below 90\,s which is negligible compared to the time scale of the transit variations (typically 30\,min for the transit ingress and egress). However the exposure time of the K2 light-curve is 29.424\,min. Consequently, for the model of the K2 data, we supersampled\footnote{We refer the reader to \citet{2010MNRAS.408.1758K} for more details regarding the need of supersampling in light-curve modelling.
} the model by a factor 10. This means that for each exposure, we computed the instantaneous value predicted by the model at 10 different times evenly distributed over the exposure and then used the average of these 10 values as the model value for the whole exposure.    

A first analysis of this system showed a linear trend in the residuals of the TRAPPIST and the September IAC80 light curves. We also noticed a more complex behavior in the November IAC80 light curves that we decomposed into two linear trends with a break point at t = 2\,457\,328.5022 HJD. Therefore we split the November IAC80 light curves into two and added 8 parameters to our model to account for these linear variations of the out-of-transit (2 per light-curve). When doing so, we used the time of the first sample ($t_{\textrm{min}}$) as the origin for the linear function: $\Delta F_\textsc{oot} + (t - t_{\textrm{min}})\ \Delta F^\prime_\textsc{oot}$). $t_{\textrm{min}}$ is equal to 2\,457\,187.753440000124,  2\,457\,269.443920060061, 2\,457\,328.353823559824, 2\,457\,328.502696809825 HJD for the TRAPPIST, the September IAC80, the first part and the second part of the November IAC80 light-curves respectively.

We re-analyzed jointly all the datasets with these 8 additional free parameters in our model. The inferred parameter values and error bars are reported in \tab{syspar}. Figure \figref{W151LCdatacomp} and \figref{W151RVdatacomp} show the photometric and radial velocity data phase folded at the best-fit ephemeris (see \tab{syspar}) with the best-fit model and residuals. The error bars displayed take into account the best-fit jitter values obtained by the Bayesian inference (see \tab{syspar}).

\paragraph{WASP-153:}~The exposure times of the WASP, Liverpool and RISE-2 data being below 40\,s, no supersampling was required for this system. The analysis didn't show any abnormal behavior. The inferred parameter values and error bars are reported in \tab{syspar} and the Figures \figref{W153LCdatacomp} and \figref{W153RVdatacomp} show the photometric and radial velocity data phase folded at the best-fit ephemeris (see \tab{syspar}) with the best-fit model and residuals. The error bars displayed take into account the best-fit jitter values obtained by the Bayesian inference (see \tab{syspar}).

\begin{figure}[!htb]
    \resizebox{\hsize}{!}{\includegraphics{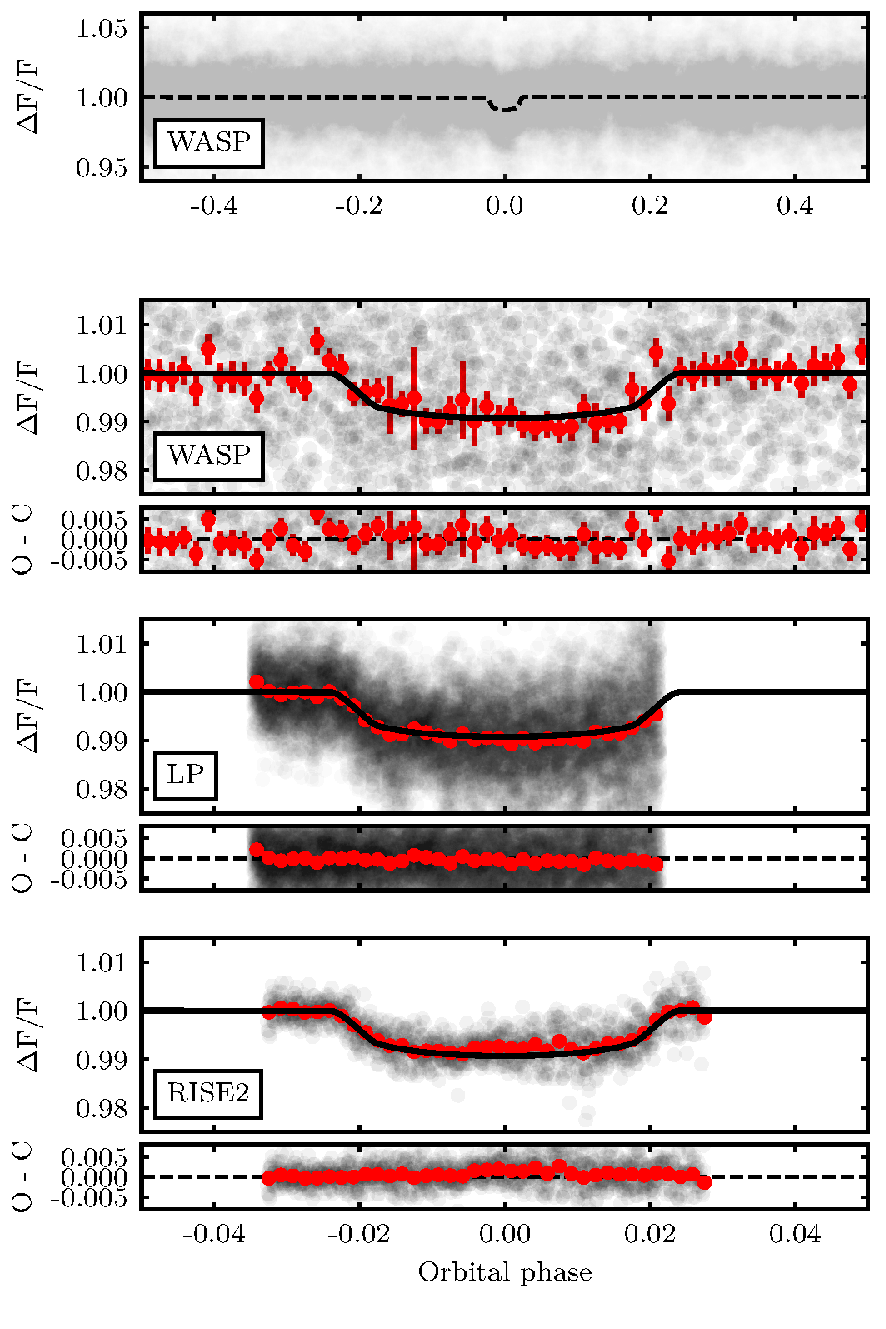}}
    \caption{\figlabel{W153LCdatacomp}Photometry of WASP-153. 
The black/grey points are the data points at the original cadence of the observations, displayed without error bars for clarity.
The red points corresponds to the same data points binned in phase with a bin width equivalent to 8 minutes. These points are represented with their associated $1\,\sigma$ error bars. The black dashed and solid lines correspond to the best-fit model at the original and binned cadence respectively.}
\end{figure}

\begin{figure}[!htb]
    \resizebox{\hsize}{!}{\includegraphics{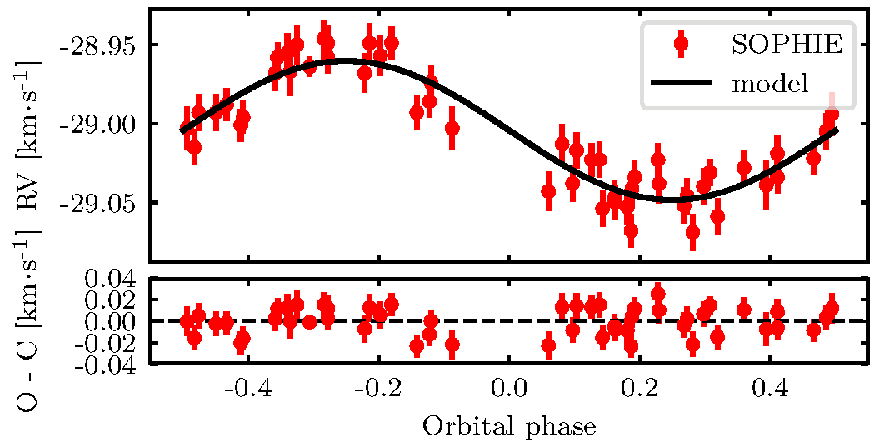}}
    \caption{\figlabel{W153RVdatacomp}Radial velocities of WASP-153. The data points are represented with their associated $1\,\sigma$ error bars.}
\end{figure}

\paragraph{WASP-156:}~The exposure times of the WASP and NITES data being both below 40\,s, no supersampling has been applied for those two datasets. The exposure time of the EulerCam data being around 80\,s and the ingress and egress for this system being relatively short ($\sim10$\,min), we decided to supersample the model by a factor 4. 
	
	A first analysis of this system showed that the 2 datasets collected with EulerCam were not compatible. The 2014 EulerCam dataset displayed a very pronounced V-shape that was not supported by the other datasets. As described in \sect{gbfupphotobs}, this dataset was identified earlier as affected by large PSF variations, and stellar counts in the non-linear regime of the EulerCam CCD. So we decided to discard it from the final analysis. We also noticed that the residuals of the 2016 EulerCam light-curve seemed to exhibit a quadratic trend and introduced 3 additional parameters to our model to account for a possible quadratic variation of the out-of-transit level. When doing so, we used the time of the first sample ($t_{\textrm{min}} = 2\,457\,700.517166$ HJD) as the origin for the quadratic function: $\Delta F_\textsc{oot} + (t - t_{\textrm{min}})\ \Delta F^\prime_\textsc{oot} + (t - t_{\textrm{min}})^2\ \Delta F'^{\prime\prime}_\textsc{oot}$). 
	
	We re-analyzed jointly all the datasets with these 3 additional free parameters in our model. The inferred parameter values and error bars are reported in \tab{syspar}. Figure \figref{W156LCdatacomp} and \figref{W156RVdatacomp} show the photometric and radial velocity data phase folded at the best-fit ephemeris (see \tab{syspar}) with the best-fit model and residuals. The error bars displayed take into account the best-fit jitter values obtained by the Bayesian inference (see \tab{syspar}).\vspace{0.2cm}	
\begin{figure}[!htb]
    \resizebox{\hsize}{!}{\includegraphics{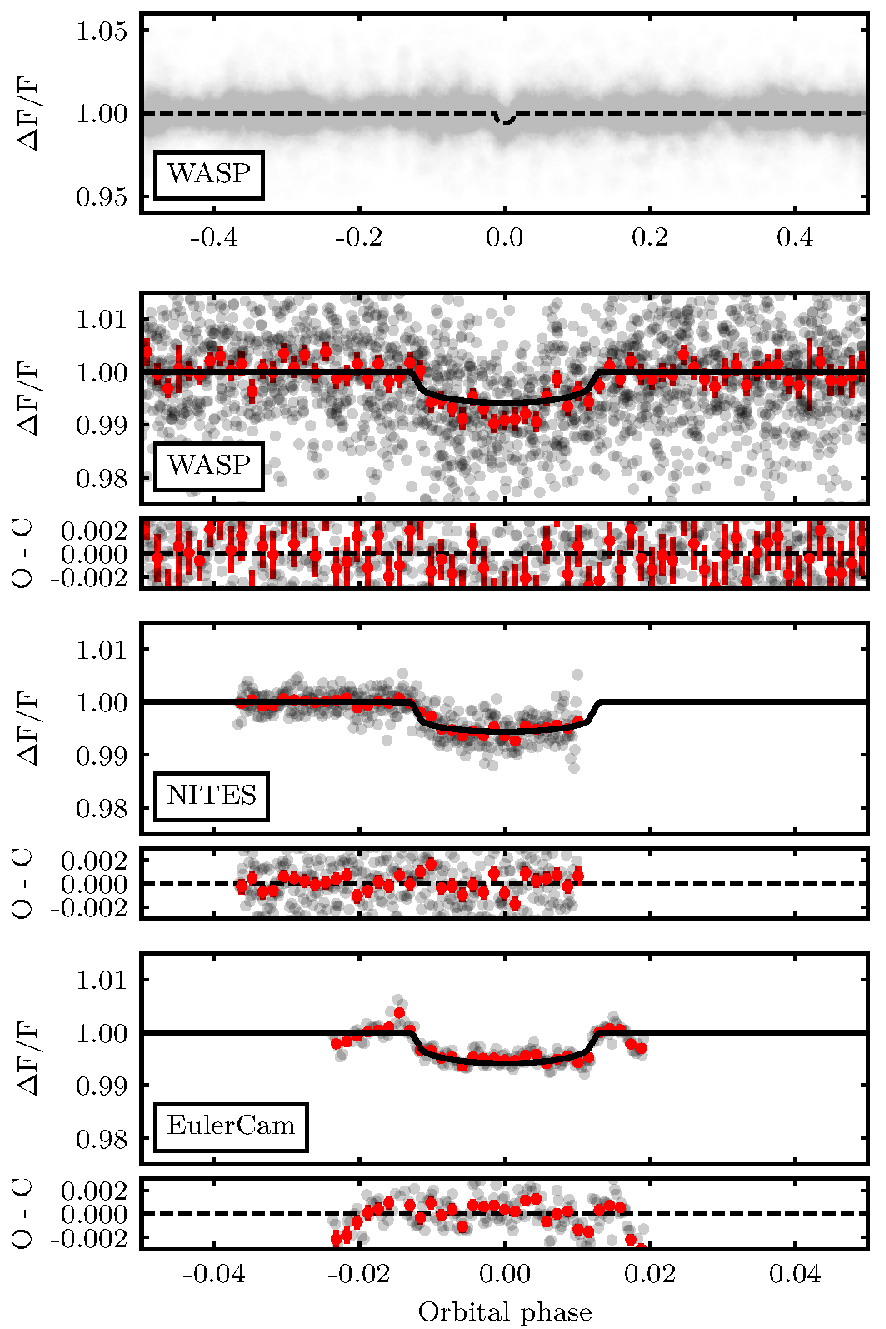}}
    \caption{\figlabel{W156LCdatacomp}Photometry of WASP-156. 
The black/grey points are the data points at the original cadence of the observations, displayed without error bars for clarity.
The red points corresponds to the same data points binned in phase with a bin width equivalent to 8 minutes. These points are represented with their associated $1\,\sigma$ error bars. The black dashed and solid lines correspond to the best-fit model at the original and binned cadence respectively.}
\end{figure}

\begin{figure}[!htb]
    \resizebox{\hsize}{!}{\includegraphics{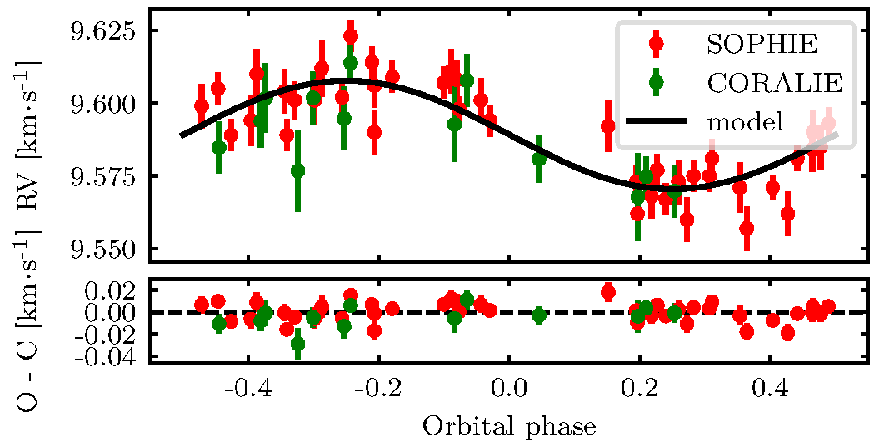}}
    \caption{\figlabel{W156RVdatacomp}Radial velocities of WASP-156. The data points are represented with their associated $1\,\sigma$ error bars.}
\end{figure}

\subsubsection{Stellar modelling}\sectlabel{stelmod}

In \sect{stelparres}, we derived stellar masses and radii from \teff, \logg\ and \feh\ using the \citet{2010A&ARv..18...67T} calibration and ages using lithium abundances and gyrochronology.
If those two age estimates seem to agree for our three systems, the Lithium constraint on the age is very weak and gyrochronology is known to sometimes contradict other ages estimators like isochronal ages \citep[e.g.][]{2016JSWSC...6A..38B, 2015MNRAS.450.1787A, 2015A&A...581A...2K, 2015A&A...577A..90M}.
Furthermore, the additional constraint brought by the stellar density inferred from the transit and a dedicated modelling of the star should result in more accurate estimates of the stellar masses and radii.
Consequently to provide a more comprehensive view of our three systems, we modelled the stars using the Fortran software \texttt{bagemass}\footnote{We used the version 1.1 available at \url{http://sourceforge.net/projects/bagemass}.}
\citep{2015A&A...575A..36M}. 

\texttt{Bagemass} relies on a grid of stellar models\footnote{\texttt{bagemass} provides several grid with different mixing length ($\alpha_{\mathrm{MLT}}$ equal 1.78 or 1.50) and different Helium-enhancement (0.0 or 0.2). For this work, we used the default values which correspond to no Helium-enhancement and $\alpha_{\mathrm{MLT}} = 1.78$. However, in \tab{fullbagemassoutput}, we present estimates of the sensitivity of the results to this assumptions.}
produced with the \textsc{garstec} stellar evolution code \citep{2008Ap&SS.316...99W}. This grid covers the mass range between $0.6$ to $2.0\,\mathrm{M_\sun}$, the initial metallicity range between $-0.75$ to $0.55\,\mathrm{dex}$ and the age range between the end of the pre-main-sequence phase up to 17.5 Gyr (or a maximum radius of $3\,\mathrm{R_\sun}$ depending on which one occurs first). In order to obtain stellar properties for any mass, metallicity and age within these ranges, and not only for the points in the grid, \texttt{bagemass} uses the cubic spline interpolation algorithm \textsc{pspline}\footnote{The \textsc{pspline} algorithm is available at \url{http://w3.pppl.gov/ntcc/PSPLINE}
}. Given measurements (values and error bars) for the $T_{\mathrm{eff}}$, $[\mathrm{Fe/H}]$ and density ($\rho_{*}$) of the star studied, it then explores this parameter space using a \textsc{mcmc} method which computes the \textit{posterior} probability as a function of mass and age.

Using the $T_{\mathrm{eff}}$ and $[\mathrm{Fe/H}]$ estimates provided by the spectral analysis  and the stellar density estimates obtained from the analysis of the transit (see \sect{sysparres} and \tab{syspar}), we obtained estimates and $68\,\%$ confidence interval error bars for the ischronal age and the mass of our three stars\footnote{The complete output table provided by \texttt{bagemass} is available in \tab{fullbagemassoutput}.
}. These values are reported in \tab{syspar}.
\fig{bagemassplot} shows the marginalized probability distribution in the Hertzsprung–Russell diagram along with the best-fit evolutionary model and isochrones for our three stars. To provide more robust error bars, the error bar provided in \tab{syspar} for the mass estimate ($M_{*} \textit{(tr. + ev. track)}$) is the square-root of the quadratic sum of the internal error and the sensitivities to the mixing length parameter and the Helium-enhancement. 
Finally, we also computed new estimates for the secondary parameters of the transit and RV analysis (see \sect{trrvanalysis}) which rely on the stellar mass and radius estimates. The most sensitive of those parameters are $R_{p}$, $M_{p}$, $\rho_{p}$, $H$ and $F_{i}$. We reported these estimates in \tab{syspar}.

The interpretation of the isochronal age estimate is the subject of \sect{agediscrepancy}, so we will now focus on the stellar mass and radius estimates.
For WASP-151, this analysis provides estimates that are compatible within one sigma with the ones obtained with the \cite{2010A&ARv..18...67T} calibration (\sect{stelparres}). However for WASP-153 and WASP-156, it's not the case. The stellar modelling predicts a significantly bigger radius for WASP-153 and a significantly lower radius for WASP-156 while the masses are compatible within one sigma (see \tab{syspar}).
This difference is mainly explained by the difference in \logg\ between the spectroscopic and transit analyses (see tr. and spec. values of \logg\ in \tab{syspar}). The comparison of \logg\ estimates from spectroscopy made by \citet{2005MSAIS...8..130S} showed that a realistic error bars for a \logg\ estimator from spectroscopy is $\sim 20\,\%$, while the one inferred from the transit density is more direct and more robust with typical uncertainties $\lesssim 5 \%$ depending on the quality of the light-curve and the photometric stellar variability.
As described in \sect{dicoveryobs}, our three stars are not particularly active. We will thus rely on the stellar mass and radius estimates obtained in this section for the rest of the paper, even if we show in \fig{massradiusdiag}, \figref{Neptuniandesert} and \tab{syspar} the estimates relying on the spectroscopic \logg\ for completeness.

\begin{figure*}[!htb]
    \centering
    \begin{minipage}{0.33\textwidth}
        \centering
        \includegraphics[width=\textwidth]{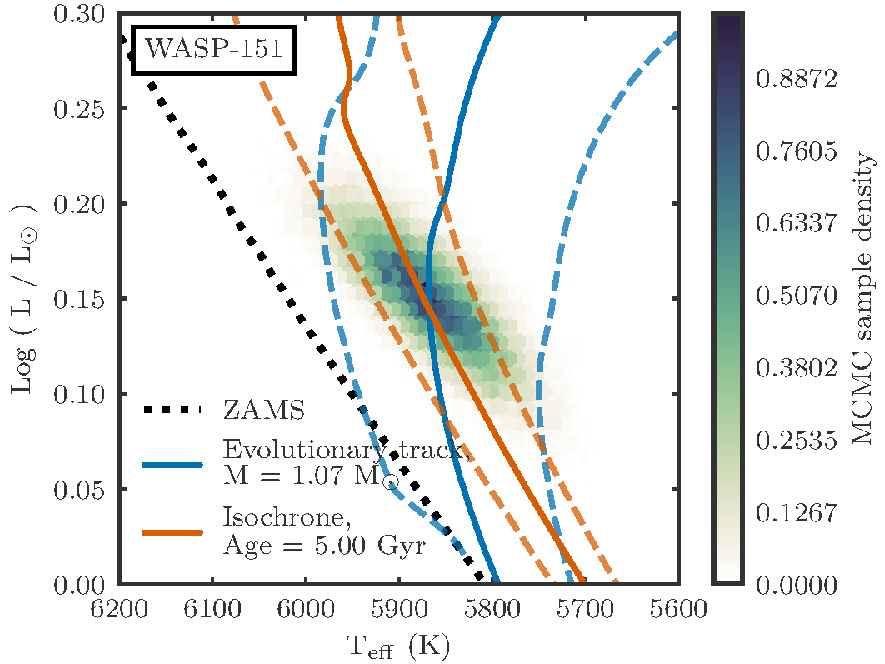}
    \end{minipage}\hfill
    \begin{minipage}{0.33\textwidth}
        \centering
        \includegraphics[width=\textwidth]{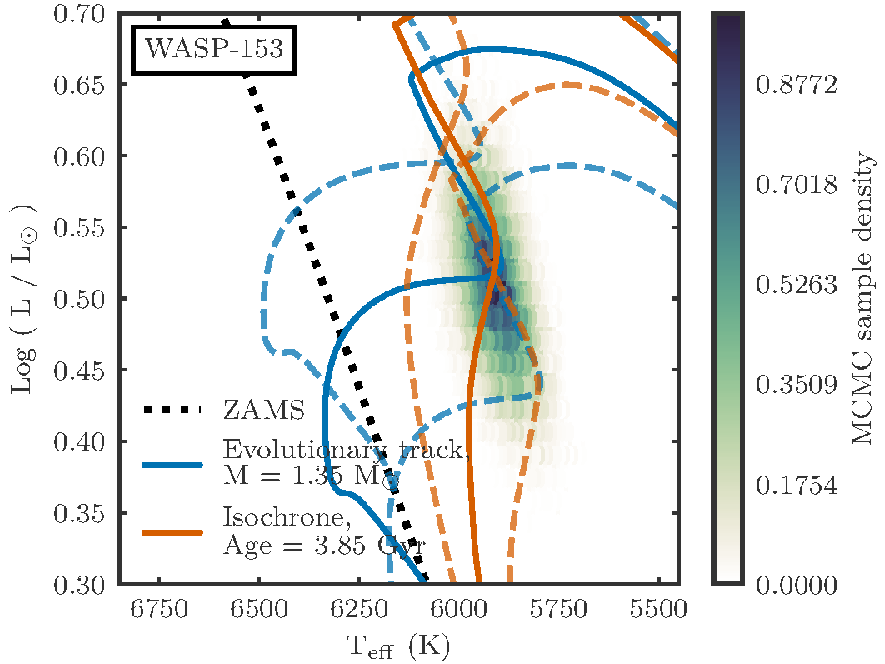}
    \end{minipage}
    \begin{minipage}{0.33\textwidth}
        \centering
        \includegraphics[width=\textwidth]{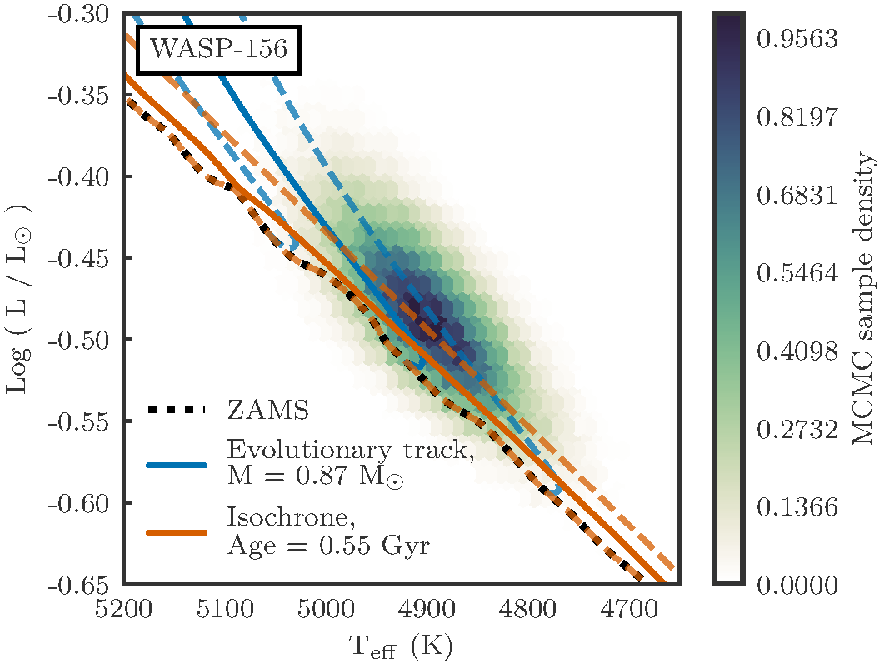}
    \end{minipage}
    \caption{\figlabel{bagemassplot}WASP-151, WASP-153 and WASP-156 marginalized posterior distribution in the Hertzsprung–Russell diagram. The dotted black line correspond to the zero-age main sequence (ZAMS) at best-fit $[\mathrm{Fe/H}]$. It's defined as the time at which the star reaches its minimum luminosity and stellar ages are measured relative to this time. The blue lines are stellar mass evolutionary tracks and the orange ones are age isochrones. For both isochrones and evolutionary tracks the solid line correspond to the best-fit model (maximum of joint likelihood distribution) and the dashed lines correspond to the two extrema of the $68\,\%$ confidence interval. For more details see \sect{stelmod} and \tab{fullbagemassoutput}.}
\end{figure*}

\begin{longtab}
\begin{longtable}{lccc} 
\caption{\tablabel{syspar}System parameters from Bayesian MCMC analysis.}\\

 & WASP-151b & WASP-153b & WASP-156b \\
\hline \\[-5pt]
\endfirsthead

\multicolumn{4}{c}%
{{\textbf{\tablename\ \thetable{}} -- continued from previous page}} \\[3pt]
 & WASP-151-b & WASP-153-b & WASP-156-b \\
\hline \\[-5pt]
\endhead

\hline
\multicolumn{4}{r}{Continued on next page\dots}\\
\multicolumn{4}{l}{See notes at the end of the table.}\\
\endfoot

\hline \hline
\multicolumn{4}{l}{{\bf Notes:}}\\
\multicolumn{4}{l}{\textit{(spec.)} indicates that the estimate has been performed using the spectroscopic data only (\sect{stelparres}).}\\
\multicolumn{4}{l}{\textit{(tr.)} indicates that the estimate has been done using transit and RV analysis only (\sect{trrvanalysis}).}\\
\multicolumn{4}{l}{\textit{(spec., tr.)} indicates that the estimate has been done using both transit and spectroscopic data (\sect{trrvanalysis}).}\\
\multicolumn{4}{l}{$M_*$ and $R_*$ \textit{(spec.)} estimates are done using the spectroscopic \teff, \logg, \feh\ and}\\
\multicolumn{4}{l}{the \cite{2010A&ARv..18...67T} calibration.}\\
\multicolumn{4}{l}{$M_*$ and $R_*$ \textit{(tr. + ev. track)} are provided by \texttt{bagemass} using $\rho_{*}$ \textit{(tr.)}, \teff\ and \feh\ \textit{(spec.)} (\sect{stelmod}).}\\
\multicolumn{4}{l}{\textit{(from spec.)} indicates that the estimate has been done using $M_*$ and $R_*$ \textit{(spec.)} estimates.}\\
\multicolumn{4}{l}{\textit{(adopted, from tr. + ev. track)} indicates that the estimate has been done using $M_*$ and $R_*$ \textit{(tr. + ev. track)} estimates}\\ 
\multicolumn{4}{l}{and that we adopted those values as final values for the system. We believe that those values are more accurate}\\
\multicolumn{4}{l}{than the one provided the spectroscopic parameters and the the \cite{2010A&ARv..18...67T} calibration.}\\
\multicolumn{4}{l}{Spectral Type are estimated from \teff\ using the table in \cite{2008oasp.book.....G}.}\\
\multicolumn{4}{l}{Abundances are relative to the solar values obtained by \cite{2009ARA&A..47..481A}.}\\
\multicolumn{4}{l}{$^{\bullet}$ indicates that the parameter is a jumping parameter in the \textsc{mcmc} analysis.}\\
\multicolumn{4}{l}{For more details on the meaning of the notations used for the parameters name, see \sect{sysparres}.}\\
\endlastfoot
 
\multicolumn{4}{l}{\textit{Planetary Parameters}} \\
\hline \\[-6pt]
$R_p$ [R$_{\textrm{Jup}}$] \textit{(adopted, from tr. + ev. track)} 							& $1.13_{-0.03}^{+0.03}$							& $1.55_{-0.08}^{+0.10}$						& $0.51_{-0.02}^{+0.02}$ \\[3pt]
$R_p$ [R$_{\textrm{Jup}}$] \textit{(from spec.)}						& $1.2_{-0.2}^{+0.2}$							& $1.1_{-0.2}^{+0.2}$						& $0.63_{-0.10}^{+0.10}$ \\[3pt]
$M_p$ [M$_{\textrm{Jup}}$] \textit{(adopted, from tr. + ev. track)}								& $0.31_{-0.03}^{+0.04}$						& $0.39_{-0.02}^{+0.02}$						& $0.128_{-0.009}^{+0.010}$ \\[3pt] 
$M_p$ [M$_{\textrm{Jup}}$] \textit{(from spec.)}								& $0.33_{-0.03}^{+0.04}$							& $0.37_{-0.02}^{+0.02}$ 						& $0.13_{-0.01}^{+0.01}$ \\[3pt]
$\rho_p$  [$\rho_{\textrm{Jup}}$] \textit{(adopted, from tr. + ev. track)}						& $0.22_{-0.02}^{+0.03}$							& $0.11_{-0.02}^{+0.02}$ 						& $1.0_{-0.1}^{+0.1}$ \\[3pt]
$\rho_p$  [$\rho_{\textrm{Jup}}$] \textit{(from spec.)}						& $0.18_{-0.06}^{+0.11}$							& $0.3_{-0.1}^{+0.2}$ 						& $0.5_{-0.2}^{+0.3}$ \\[3pt]
$T_{\textrm{eq}}$ [K]										& $1.29\,10^{3}\phantom{\,}_{-1\,10^{1}}^{+2\,10^{1}}$	&
$1.70\,10^{3}\phantom{\,}_{-4\,10^{1}}^{+4\,10^{1}}$	& $9.7\,10^{2}\phantom{\,}_{-2\,10^{1}}^{+3\,10^{1}}$\\[3pt]
${P}^{\ \bullet}$\ [days]									& $4.533471_{-4\,10^{-6}}^{+4\,10^{-6}}$				& $3.332609_{-2\,10^{-6}}^{+2\,10^{-6}}$			& $3.836169_{-3\,10^{-6}}^{+3\,10^{-6}}$ \\[3pt]
${t_c}^{\bullet}$\ [BJD - 2\,400\,000]							& $57741.0081_{-2\,10^{-4}}^{+1\,10^{-4}}$			& $53142.542_{-0.003}^{+0.003}$				& $54677.707_{-0.002}^{+0.002}$ \\[3pt]
$a$ [AU]												& $0.055_{-0.001}^{+0.001}$					& $0.048_{-0.001}^{+0.001}$				& $0.0453_{-0.0009}^{+0.0009}$ \\[3pt]
$e$ 													& $< 0.003$							& $< 0.009$								& $< 0.007$ \\[3pt]
$\omega_*$ [$^\circ$]									& $-5\,10^{1}\phantom{\,}_{-3\,10^{1}}^{+1.2\,10^{2}}$	& unconstrained							& unconstrained \\[3pt]
$i$ [$^\circ$]											& $89.2_{-0.6}^{+0.6}$							& $84.1_{-0.7}^{+0.7}$						& $89.1_{-0.9}^{+0.6}$ \\[3pt]
${\sqrt{e}\cos \omega_*}^{\bullet}$ 							& $0.000_{-9\,10^{-3}}^{+1.0\,10^{-2}}$				& $-0.00_{-0.05}^{+0.05}$						& $-0.01_{-0.10}^{+0.04}$ \\[3pt]
${\sqrt{e}\sin \omega_*}^{\bullet}$ 							& $0.00_{-0.04}^{+0.03}$							& $-0.00_{-0.05}^{+0.05}$						& $0.00_{-0.05}^{+0.05}$ \\[3pt]
${\cos i}^{\ \bullet}$ 										& $0.015_{-0.010}^{+0.009}$						& $0.10_{-0.01}^{+0.01}$						& $0.02_{-0.01}^{+0.02}$ \\[3pt]
${a / R_*}^{\ \bullet}$ 									& $10.34_{-0.19}^{+0.11}$							& $6.0_{-0.2}^{+0.3}$ 						& $12.8_{-0.7}^{+0.3}$ \\[3pt]
${R_p / R_*}^{\ \bullet} $ [\%]								& $10.11_{-0.03}^{+0.05}$							& $9.2_{-0.1}^{+0.1}$					& $6.85_{-0.08}^{+0.12}$ \\[3pt]
$\Delta F/ F$ [\%]										& $1.021_{-0.007}^{+0.01}$						& $0.85_{-0.02}^{+0.02}$						& $0.47_{-0.01}^{+0.02}$ \\[3pt]
$D14$ [h]												& $3.66_{-0.01}^{+0.02}$							& $3.84_{-0.05}^{+0.05}$						& $2.41_{-0.03}^{+0.04}$ \\[3pt]
$D23$ [h]	 											& $2.97_{-0.02}^{+0.01}$							& $2.83_{-0.06}^{+0.07}$						& $2.08_{-0.03}^{+0.03}$\\[3pt]
${K}^{\ \bullet}$ [\ms]										& $37_{-3}^{+4}$								& $44_{-2}^{+2}$							& $19_{-1}^{+1}$ \\[3pt]
$\tau_{\textrm{circ}}$ [Gyr] 								& $0.03_{-0.003}^{+0.004}$							& $0.0022_{-0.0006}^{+0.0006}$				& $0.26_{-0.05}^{+0.06}$ \\[3pt]
$H$ [km] \textit{(adopted, from tr. + ev. track)}												& $8.0\,10^{2}\phantom{\,}_{-8\,10^{1}}^{+9\,10^{1}}$		& $1.6\,10^{3}\phantom{\,}_{-2\,10^{2}}^{+2\,10^{2}}$	& $3.0\,10^{2}\phantom{\,}_{-3\,10^{1}}^{+4\,10^{1}}$ \\[3pt] 
$H$ [km] \textit{(from spec.)}												& $9.0\,10^{2}\phantom{\,}_{-3\,10^{2}}^{+3\,10^{2}}$	& $8\,10^{2}\phantom{\,}_{-2\,10^{2}}^{+3\,10^{2}}$	& $5\,10^{2}\phantom{\,}_{-1\,10^{2}}^{+2\,10^{2}}$ \\[3pt]
$F_{i}$ [$F_{i, \oplus}$] \textit{(adopted, from tr. + ev. track)}									& $4.6\,10^{2}\phantom{\,}_{-2\,10^{1}}^{+2\,10^{1}}$		& 
$1.4\,10^{3}\phantom{\,}_{-1\,10^{2}}^{+2\,10^{2}}$ &$1.5\,10^{2}\phantom{\,}_{-1\,10^{1}}^{+1\,10^{1}}$ \\[3pt] 
$F_{i}$ [$F_{i, \oplus}$] \textit{(from spec.)}									& $5\,10^{2}\phantom{\,}_{-1\,10^{2}}^{+2\,10^{2}}$		& 
$7\,10^{2}\phantom{\,}_{-2\,10^{2}}^{+3\,10^{2}}$ &$2\,10^{2}\phantom{\,}_{-7\,10^{1}}^{+8\,10^{1}}$ \\[3pt]

\\[-3pt]
\multicolumn{4}{l}{\textit{Stellar Parameters}} \\
\hline \\[-5pt]
RA  [hours:minutes:sec]									& 23:16:15.22									& 18:37:02.97 						& 02:11:07.61 \\[3pt]
DEC [degrees minutes sec]	 							& 00 18 24.5									& 40 01 07.4 						& 02 25 04.8 \\[3pt]
Sp. Type \textit{(spec.)}    									& G1									& G0 						& K3 \\[3pt]
V mag												& 12.9								& 12.8 						& 11.6 \\[3pt]
J mag												& 11.5								& 11.4 						& 9.9 \\[3pt]
V - K													& 1.7									& 1.7 						& 2.2 \\[3pt]
$M_*$ [M$_{\sun}$] \textit{(adopted, tr. + ev. track.)}      						& $1.077 \pm 0.081$						& $1.336 \pm 0.086$				& $0.842 \pm 0.052$ \\[3pt]
$M_*$ [M$_{\sun}$] \textit{(spec.)}      						& 1.14 $\pm$ 0.09 						& 1.20 $\pm$ 0.09 				& 0.87 $\pm$ 0.07 \\[3pt]
$R_*$  [R$_{\sun}$] \textit{(adopted, tr. + ev. track.)}       						& $1.14^{+0.03}_{-0.03}$						& $1.73^{+0.10}_{-0.09}$ 				& $0.76^{+0.03}_{-0.03}$ \\[3pt]
$R_*$  [R$_{\sun}$] \textit{(spec.)}       						& 1.24 $\pm$ 0.18						& 1.18 $\pm$ 0.20 				& 0.95 $\pm$ 0.15 \\[3pt]
Age [Gyr] \textit{(adopted, Isochrone)}					& $5.1^{+1.3}_{-1.3}$					& $4.0^{+0.8}_{-0.8}$			& $6.4^{+4.0}_{-4.0}$ \\[3pt]
Age [Gyr] \textit{(Gyrochronology)}									& $1.80^{+2.03}_{-1.00}$					& $1.21^{+1.19}_{-0.60}$			& $0.58^{+0.51}_{-0.31}$ \\[3pt]
$\rho_*$ [$\rho_\sun$] \textit{(tr.)} 							& $0.72_{-0.04}^{+0.02}$					& $0.26_{-0.03}^{+0.04}$ 			& $1.9_{-0.3}^{+0.1}$\\[3pt]
\teff\ [K]  \textit{(spec.)}  									& 5871 $\pm$ 57 						& 5914 $\pm$ 64 				& 4910 $\pm$ 61\\[3pt]
\logg\ \textit{(spec.)}   									& 4.30 $\pm$ 0.11 						& 4.36 $\pm$ 0.13 				& 4.40 $\pm$ 0.12\\[3pt]
\logg\ \textit{(adopted, tr.)} 										& $4.35_{-0.03}^{+0.02}$					& $4.10_{-0.06}^{+0.06}$		& $4.60_{-0.07}^{+0.04}$\\[3pt]
{[Fe/H]} [dex] \textit{(spec.)}   								& 0.10 $\pm$ 0.10 						& 0.34 $\pm$ 0.11 				& 0.24 $\pm$ 0.12\\[3pt]
\vsini\ [\kms] \textit{(spec.)} 								& 4.25 $\pm$ 0.90 						& 5.19 $\pm$ 0.95 				& 3.80 $\pm$ 0.91\\[3pt]
\mactrb\ [\kms] \textit{(spec.)}   								& 3.73 $\pm$ 0.73 						& 3.73 $\pm$ 0.73  				& 2.77 $\pm$ 0.73\\[3pt]
\mictrb\ [\kms] \textit{(spec.)}     								& 0.32 $\pm$ 0.10 						& 0.59 $\pm$ 0.06 				& \\[3pt]
$\log A$(Li) \textit{(spec.)}    								& 1.73 $\pm$ 0.05 						& 2.77 $\pm$ 0.05 				& $<$0.19 $\pm$ 0.08\\[3pt]
Distance [pc] \textit{(spec.)}    								& 480 $\pm$ 75						& 430 $\pm$ 35 				& 140 $\pm$ 25 pc \\[3pt]
${v0_{\textrm{SOPHIE}}}^{\bullet}$ [\kms] 	\textit{(tr.)}				& $-12.369_{-0.002}^{+0.002}$				& $-29.004_{-0.001}^{+0.002}$		& $9.5891_{-9\,10^{-4}}^{+9\,10^{-4}}$ \\[3pt]
${u_{\textrm{Johson}\,R}}^{\bullet}$ \textit{(spec., tr.)}			& $0.478_{-0.002}^{+0.002}$				& $0.486_{-0.002}^{+0.002}$ 		& $0.591_{-0.003}^{+0.003}$ \\[3pt]
${v_{\textrm{Johson}\,R}}^{\bullet}$ \textit{(spec., tr.)}				& $0.129_{-0.005}^{+0.005}$				& $0.126_{-0.005}^{+0.005}$ 		& $0.082_{-0.008}^{+0.008}$ \\[3pt]
${u_{\textrm{Sloan}\,z}}^{\bullet}$ \textit{(spec., tr.)} 				& $0.341_{-0.001}^{+0.001}$				& 							&\\[3pt]
${v_{\textrm{Sloan}\,z}}^{\bullet}$ \textit{(spec., tr.)} 				& $0.127_{-0.004}^{+0.004}$				& 							&\\[3pt]
${u_{\textrm{Kp}}}^{\bullet}$ \textit{(spec., tr.)} 					& $0.549_{-0.003}^{+0.003}$				& 							&\\[3pt]
${v_{\textrm{Kp}}}^{\bullet}$ \textit{(spec., tr.)} 					& $0.114_{-0.005}^{+0.005}$				& 							&\\[3pt]
${u_{\textrm{NGTS}}}^{\bullet}$	\textit{(spec., tr.)}				& $0.487_{-0.002}^{+0.002}$				& 							&\\[3pt]
${v_{\textrm{NGTS}}}^{\bullet}$	\textit{(spec., tr.)}				& $0.131_{-0.005}^{+0.004}$				& 							&\\[3pt]
${u_{\textrm{V+R}}}^{\bullet}$ \textit{(spec., tr.)}				& 									& $0.549_{-0.002}^{+0.002}$							&  \\[3pt]
${v_{\textrm{V+R}}}^{\bullet}$ \textit{(spec., tr.)}				& 									& $0.118_{-0.006}^{+0.006}$ 							&  \\[3pt]
${u_{\textrm{Johson}\,I}}^{\bullet}$ \textit{(spec., tr.)}				& 									&  							& $0.461_{-0.002}^{+0.002}$ \\[3pt]
${v_{\textrm{Johson}\,I}}^{\bullet}$ \textit{(spec., tr.)}				& 									&  							& $0.0995_{-0.0074}^{+0.0066}$ \\[3pt]
${u_{\textrm{Gunn}\,r}}^{\bullet}$ \textit{(spec., tr.)}				& 									&  							& $0.461_{-0.002}^{+0.002}$ \\[3pt]
${v_{\textrm{Gunn}\,r}}^{\bullet}$ \textit{(spec., tr.)}				& 									&  							& $0.103_{-0.007}^{+0.007}$ \\[3pt]
\\[-5pt]
\multicolumn{4}{l}{\textit{Instruments Parameters}} \\
\hline \\[-5pt]
${\Delta\textrm{RV}_{\textrm{CORALIE/SOPHIE}}}^{\bullet}$ [\kms]			& $0.055_{-0.009}^{+0.008}$						& 						& $0.043_{-0.002}^{+0.002}$\\[3pt]
${\ln f_{\sigma \textrm{SOPHIE}}}^{\bullet}$							& $0.03_{-0.05}^{+0.04}$							& $0.04_{-0.04}^{+0.04}$		& $0.08_{-0.04}^{+0.04}$ \\[3pt]
${\ln f_{\sigma \textrm{CORALIE}}}^{\bullet}$							& $-0.01_{-0.06}^{+0.06}$							& 						& $-0.01_{-0.05}^{+0.05}$ \\[3pt]
${\ln f_{\sigma \textrm{K2}}}^{\bullet}$								& $1.71_{-0.02}^{+0.02}$							&  						&\\[3pt]
${\ln f_{\sigma \textrm{EulerCam}}}^{\bullet}$							& $0.24_{-0.03}^{+0.03}$							& 						& $0.43_{-0.03}^{+0.03}$ \\[3pt]
${\ln f_{\sigma \textrm{TRAPPIST}}}^{\bullet}$							& $0.01_{-0.02}^{+0.03}$							&  						&\\[3pt]
${\ln f_{\sigma \textrm{WASP}}}^{\bullet}$								& $0.112_{-0.009}^{+0.008}$						&  $-0.056_{-0.008}^{+0.008}$	& $0.31_{-0.01}^{+0.01}$ \\[3pt]
${\ln f_{\sigma \textrm{Liverpool}}}^{\bullet}$							& 											&  $-1.601_{-0.006}^{+0.006}$	& \\[3pt]
${\ln f_{\sigma \textrm{RISE2}}^{\bullet}}$								& 											& $-0.81_{-0.02}^{+0.02}$						& \\[3pt]
${\ln f_{\sigma \textrm{NITES}}}^{\bullet}$								& 											&  						& $0.39_{-0.02}^{+0.03}$\\[3pt]
${\ln f_{\sigma \textrm{IAC80}}}^{\bullet}$								& $-0.32_{-0.05}^{+0.04}$	&  						&\\[3pt]
																		& $-0.26_{-0.05}^{+0.05}$							&  						&\\[3pt]
																		& $-0.14_{-0.05}^{+0.05}$							&  						&\\[3pt]
${\Delta F_{\textsc{oot}, \textrm{IAC80}}}^{\bullet}$					& $0.0042_{-4\,10^{-4}}^{+4\,10^{-4}}$				&  						&\\[3pt]
															& $-7\,10^{-4}\phantom{\,}_{-6\,10^{-4}}^{+6\,10^{-4}}$	&  						&\\[3pt]
															& $-1\,10^{-4}\phantom{\,}_{-8\,10^{-4}}^{+8\,10^{-4}}$	&  						&\\[3pt]
${\Delta F_{\textsc{oot}, \textrm{IAC80}}^\prime}^{\bullet}$ [day$^{-1}$]	& $-0.024_{-0.003}^{+0.004}$						&  						&\\[3pt]
															& $-0.008_{-0.007}^{+0.007}$						&  						&\\[3pt]
															& $0.02_{-0.02}^{+0.02}$							&  						&\\[3pt]
${\Delta F_{\textsc{oot}, \textrm{TRAPPIST}}}^{\bullet}$							& $-2\,10^{-6}\phantom{\,}_{-3\,10^{-4}}^{+3\,10^{-4}}$	&  						&\\[3pt]	
${\Delta F_{\textsc{oot}, \textrm{TRAPPIST}}^\prime}^{\bullet}$	 [day$^{-1}$]			& $0.008_{-0.004}^{+0.004}$						&  						&\\[3pt]
${\Delta F_{\textsc{oot}, \textrm{EulerCam}}}^{\bullet}$							& 											& 						& $7\,10^{-6}\phantom{\,}_{-1\,10^{-4}}^{+1\,10^{-4}}$\\[3pt]	
${\Delta F_{\textsc{oot}, \textrm{EulerCam}}^\prime}^{\bullet}$ [day$^{-1}$]			& 											& 						& $-5\,10^{-4}\phantom{\,}_{-9\,10^{-4}}^{+9\,10^{-4}}$\\[3pt]
${\Delta F_{\textsc{oot}, \textrm{EulerCam}}^{\prime\prime}}^{\bullet}$ [day$^{-2}$]		& 											& 						& $-0.001_{-0.001}^{+0.001}$\\[3pt]
\end{longtable}
\end{longtab}

\section{Discussion and conclusion}\sectlabel{discconclusion}

\begin{figure*}[!htb]
    \resizebox{\hsize}{!}{\includegraphics{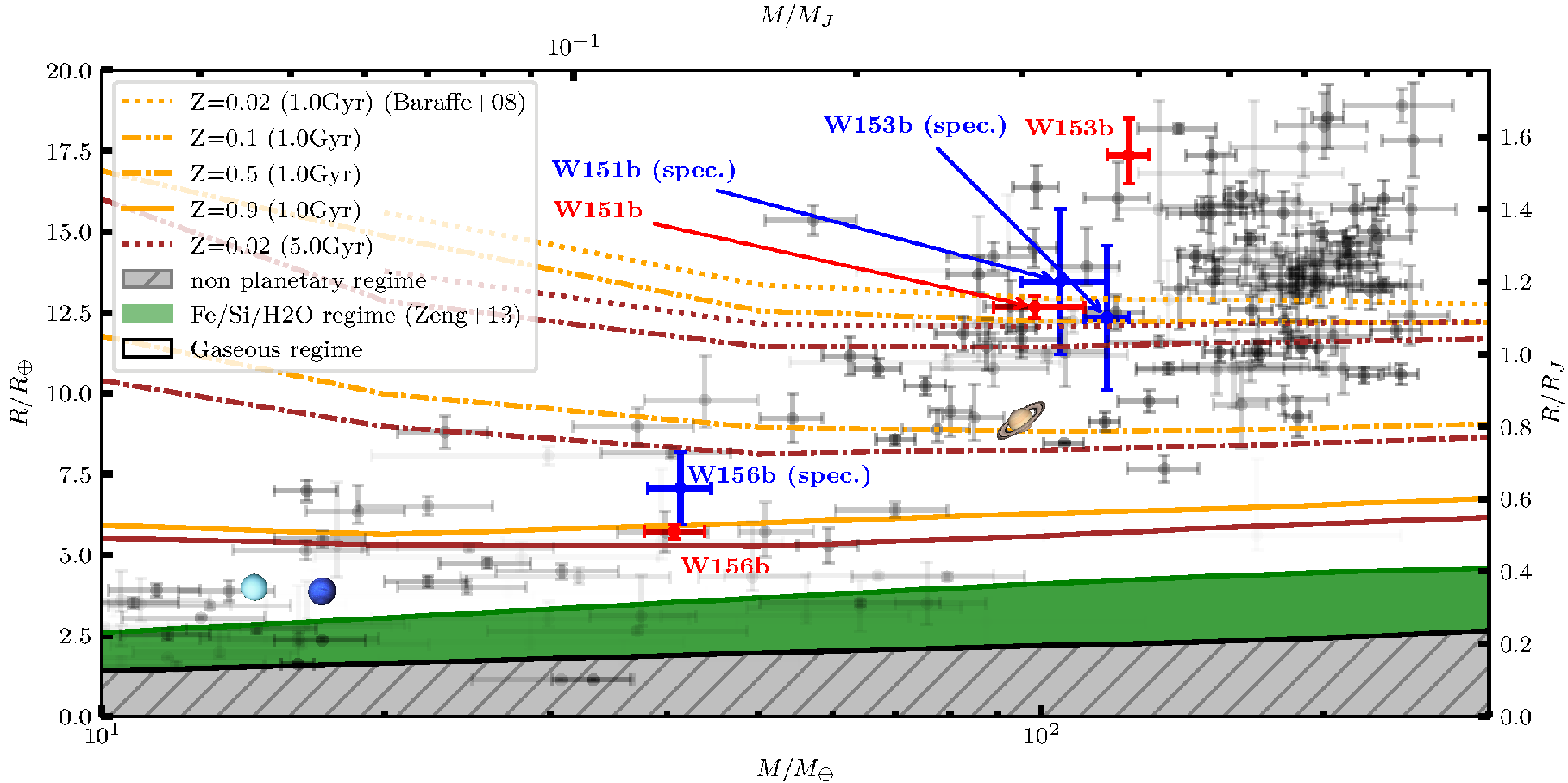}}
    \caption{\figlabel{massradiusdiag}WASP-151b, WASP-153b and WASP-156b in the mass-radius diagram. The black points with $1\,\sigma$ error bars are the known confirmed planets according to exoplanet.eu \citep{2011A&A...532A..79S}. Their transparency reflect the relative precision on their mass and radius. The better the parameter of a planet are constrained the more opaque the point is. The red and blue points with $1\,\sigma$ errors bars are the planets announced by this paper. For the red points the mass and radius estimates rely on stellar parameters obtained via evolutionary tracks and the stellar density inferred from the transit (see \sect{stelmod}), while for the blue points, they rely on purely spectroscopic stellar parameters (see \sect{stelparres}). The solar system planets (Saturn, Uranus and Neptune) have also been reported in this diagram for reference.
The two shaded areas at the bottom of the graph define the non-planetary regime (gray striped) and the rocky-water world regime (green) as defined by \citep{2013PASP..125..227Z}. Consequently the rest of the diagram represents the gaseous-ice giant regime.
The solid, dotted and dashed lines represent the mass-radius relations for gaseous planets of different age and different heavy element mass fraction ($Z$) as described by \citep{2008A&A...482..315B}. The type of line (solid or dashed or dotted) represents the heavy element mass fraction and the color of the line represents the age. 
These models have been used to constrain the nature and composition of WASP-151b, WASP-153b and WASP-156b, see Sections \sectref{compoW151&W153} and \sectref{compoW156}.}
\end{figure*}

\tab{syspar} give us an exhaustive picture of these 3 systems and allows us to put them in context.

WASP-151b and WASP-153b are relatively similar. Their masses of $0.31$ and $0.39\,\mathrm{M_{Jup}}$ and semi-major axes of  $0.056\,\mathrm{AU}$ and $0.048\,\mathrm{AU}$ respectively indicate two Saturn-size objects around early G type stars of V magnitude $\sim 12.8$. WASP-156b's radius of $0.51\,R_{\textrm{Jup}}$ suggests a Super-Neptune\footnote{\citet{2015ApJ...813..111B} defined the class of Super-Neptunes as the planets whose mass lies between $0.054\,\mathrm{M_{Jup}}$ (the mass of Neptune) and $0.18\,\mathrm{M_{Jup}}$ (halfway between the mass of Neptune and Saturn).}
and makes it the smallest planet ever detected by WASP.
Its mass of $0.128\,M_{\textrm{Jup}}$ is also the 3rd lightest detected by WASP after WASP-139b \citep[][]{2017MNRAS.465.3693H} and WASP-107b \citep[][]{2017arXiv170103776A}. Also interesting is the fact that WASP-156 is a bright ($\mathrm{magV} = 11.6$) K type star.

In the following two sections, we compared the position of our three planets in the mass-radius diagram with the isochrones of \citet[][]{2008A&A...482..315B} to constrain their composition. \citet[][]{2008A&A...482..315B} provide two types of models, one without irradiation and one with the irradiation received at $0.045$\,AU from the Sun. Given the semi-major axes of our planets, the latter is the most suited to this study and is the one we used in \fig{massradiusdiag}.
We refer readers interested in the details of these models to the associated publication. 

In the third section, we discuss the age estimates of those three systems. More specifically, we address the apparent discrepancy between the gyrochronological and isochronal ages and the possible insight that it provides regarding the migration mechanism of the planets in these systems. 
Finally, the fourth section is devoted to the impact of these three planets on our understanding of the Neptunian desert \citep{2016A&A...589A..75M}.

\subsection{Two hot Saturns: WASP-151b and WASP-153b}\sectlabel{compoW151&W153}

WASP-151b and WASP-153b's position in the mass-radius diagram indicate two low density gaseous planets (see \fig{massradiusdiag}). Their masses are close to the one of Saturn but their radii are significantly bigger, especially for WASP-153b.
Relying on its isochronal age and its relative position compared to the isochrones of \citet[][]{2008A&A...482..315B}, WASP-151b should have a heavy-element mass fraction slightly smaller than $2\,\%$. Similarly, WASP-153b's heavy-element mass fraction should be significantly smaller than $2\,\%$. Knowing that WASP-151 has a metallicity compatible with the one of the Sun, that WASP-153 is super-metallic ($\feh = 0.34 \pm 0.11$ dex) and that the Sun's heavy element mass fraction is close to $2\,\%$ \citep[e.g.][]{2008A&A...482..315B}, these heavy element mass fractions inferior to $2\,\%$ are unlikely. Consequently, WASP-151b appears to be slightly more bloated than the models predict and WASP-153b exhibits a significant radius anomaly.
This interpretation is, of course, dependant on the accuracy of our mass, radius and age estimates. As shown in \fig{massradiusdiag}, if we rely on the planetary radius inferred from the purely spectroscopic stellar parameters (\sect{stelparres}), WASP-151b and WASP-153b are compatible within one sigma with the model of \citet[][]{2008A&A...482..315B}. However as discussed in \sect{stelmod}, these estimates appear less precise and less accurate than the ones above, which rely on the stellar density inferred from the transit and stellar models.

Given the relatively high incident flux received by these two planets ($460\,\mathrm{F_{i, \oplus}}$ for WASP-151b and $1400\,\mathrm{F_{i, \oplus}}$ for WASP-153b), the radius anomalies that they exhibit was expected. Indeed, it is in agreement with the empirical thresholds defined by \citet[][]{2011ApJ...736L..29M} and \citet[][]{2016ApJ...818....4L} for an abnormally inflated radius: $R > 1.2\ \mathrm{R_{Jup}}$ and $F_i > 2\,10^{8}\ \mathrm{erg.s^{-1}.cm^{-2}} \sim 150\ \mathrm{F_{i, \oplus}}$. WASP-153b exceeds significantly both thresholds and WASP-151b exceeds the incident flux threshold, but is slightly below the radius threshold.

\subsection{A warm Super-Neptune: WASP-156b}\sectlabel{compoW156}

WASP-156b's position in the mass-radius diagram suggests a composition significantly different from the ones of WASP-151b and WASP-153b. \citet[][]{2008A&A...482..315B} models indicate a high heavy element mass fraction around 90 \%, in agreement with the one of Neptune and Uranus \citep[][]{2017arXiv170509320H}, depicting WASP-156b as a warm Super-Neptune.
Super-Neptunes with precise determination of the mass and radius (better than 15 \%) are relatively rare since only 9 of these objects are known at the moment: Kepler-9c \citep[][]{2011ApJ...727...24T}, Kepler-35b \citep[][]{2012Natur.481..475W}, Kepler-101b \citep[][]{2014A&A...572A...2B}, HATS-7b \citep[][]{2015ApJ...813..111B}, HATS-8b \citep[][]{2015AJ....150...49B}, WASP-107b \citep{2017arXiv170103776A}, WASP-127b \citep{2017A&A...599A...3L}, WASP-139b \citep{2017MNRAS.465.3693H} and  WASP-156b. 
Amongst this class of planets, WASP-156b, as a warm ($T_{\mathrm{eq}} = 970\,\mathrm{K}$) and dense ($\rho_{\mathrm{p}} = 1.0\,\mathrm{\rho_{Jup}}$) Super-Neptune, is particularly interesting to investigate the gaseous to ice giant transition as described by \citet[][]{2017arXiv170103776A} and \citet[][]{2015ApJ...813..111B}. WASP-156 is also currently the brightest Super-Neptune host star, with a V magnitude of 11.6, making it a target of prime interest for future atmospheric characterisation.

\subsection{Discrepancy between the ages estimators, an insight on migration mechanisms ?}\sectlabel{agediscrepancy}

In Sections \sectref{stelparres} and \sectref{stelmod}, we derived ages for our three stars with Lithium abundance, gyrochronology and isochrone fitting. These results are reported in \tab{ages}. The tendency that arises from this table is that our stars tend to have isochronological ages that are significantly higher than their gyrochronological ages. This tendency, limited here to three cases, has already been observed by \citet{2015A&A...577A..90M} for a broader sample of 28 transiting exoplanets where at least half of the sample exhibits this discrepancy. Interestingly for more than $80\,\%$ of the stars in this sample, and for our three stars, the planetary companion is a short period ($< 5\,\mathrm{days}$) giant planet.

Discrepancies between gyrochronological and isochronological ages have been reported by several studies and not only in the context of planet host stars, see for example \citet{2015MNRAS.450.1787A, 2015A&A...581A...2K, 2016JSWSC...6A..38B}. \citet{2015A&A...577A..90M} found that gyrochronological age estimates were significantly lower than the isochronological ones for about half of their sample of planetary hosts. \citet{2015A&A...581A...2K} reached a similar conclusion from a galactic field stars sample. Finally \citet{2016JSWSC...6A..38B} and \citet{2015MNRAS.450.1787A} brought to light inconsistencies in the gyrochronological age estimator when applied to different samples. This problem is thus complex and has multiple facets. Consequently, it will not be solved solely by the 3 stars discussed in this paper. However they can give us insights regarding the specific question of the underestimation provided by the gyrochronological age estimator observed for a fraction of the short period planet host stars population.

To explain the hot giant planet population, the core-accretion scenario requires a mechanism to migrate these planets from their formation location, beyond the ice line, to the vicinity of their parent star. There is currently two mechanisms debated in the literature for this migration: disk driven migration \citep[e.g.][]{1996Natur.380..606L,1997Icar..126..261W} and high eccentricity migration \citep[e.g.][]{1996Sci...274..954R,2007ApJ...669.1298F}. The main observational arguments to favour one over the other are: Spin-Orbit misalignment \citep[e.g.][]{2012ApJ...754L..36N}, stellar metallicity \citep[e.g.][]{2013ApJ...767L..24D}, the presence of additional companions \citep[e.g.][]{2016ApJ...825...62S} and the Roche separation \citep[e.g.][]{2017arXiv170309711N}. 

Under the light of \tab{ages} and the study performed by \citet{2015A&A...577A..90M}, we suggest that a gyrochronological age significantly smaller than the isochronal one could be an evidence to identify the mechanism responsible for the migration of giant planets. A gyrochronological ages significantly lower than the isochronological one might indeed be explained by the important transfer of angular momentum from the giant planet to the star during the tidal circularisation of the planet's orbit involved in high eccentricity migration. On the contrary, disk driven migration implies an exchange of angular momentum between the planet and the disk and cannot directly explain an increase of the stellar rotation. Furthermore, contrary to disk driven migration, high eccentricity migration is not bounded to the short protoplanetary disk lifetime and can occur at an older stage of the system amplifying even more the discrepancy between the two age estimates. If this hypothesis is confirmed for stars hosting short period planets, a gyrochronological age significantly smaller the isochronal age (e.g. the three host star presented in this paper) would indicate that the planet migrated through high-eccentricity migration while a gyrochronological age compatible with the isochronal one \citep[e.g. WASP-33][]{2010MNRAS.407..507C} would suggest a disk driven migration (or an \textit{in-situ} formation).

Obviously, a more thorough analysis is necessary to investigate all the possible implications behind this hypothesis. Such an analysis is out of the scope of this paper but we think that this hypothesis is worth investigating. In this context, a search for long period companions that might have triggered the high eccentricity migration or an independent age estimate through asterosismology with TESS \citep{2016ApJ...830..138C} or Plato \citep{2014ExA....38..249R} would be particularly interesting.

\begin{table}[htb]
\caption{\tablabel{ages}Age estimates of WASP-151, WASP-153 and WASP-156. Iso. stands for isochronal age, Gyro. for gyrochronological age and Li for the age constraint based and Lithium abundance.}
\begin{tabular}{llll}
\hline
 Star & Iso. [Gyr] & Gyro. [Gyr] & Li \\ 
\hline
\hline
WASP-151 	& $5.13_{-1.33}^{+1.33}$	& $1.80_{-1.00}^{+2.03}$	& several Gyr\\[3pt]
WASP-153 	& $4.00_{-0.77}^{+0.77}$ 	& $1.21_{-0.60}^{+1.19}$ 	& several Myr\\[3pt]
WASP-156 	& $6.50_{-4.03}^{+4.03}$ 	& $0.58_{-0.31}^{+0.51}$ 	& $\gtrsim 500$ Myr\\[3pt]
\hline
\end{tabular}
\\
\textbf{Note}: The isochronal age estimates in this table are obtained using the mean value of the marginalized \textit{posterior} distribution of the age. For WASP-151 and WASP-153, these are compatible with the maximum-likelihood estimate. However for WASP-156, it is not the case since the latter give an age of $0.5$ Gyr (see \tab{fullbagemassoutput}).
\end{table}

\subsection{Three planets at the border of the Neptunian desert}\sectlabel{Neptuniandesert}

As described in the introduction, \citet{2016A&A...589A..75M} studied the distribution of the planet population in the orbital period, mass and radius domain and reported the lower and upper mass and radius boundaries of the short period Neptunian desert. \fig{Neptuniandesert} shows that WASP-151b and WASP-153b lie near the upper boundaries of the desert, while WASP-156b stands well inside it. The authors mentioned that the period limit of the desert was not well constrained, however they also indicated that these borders delineate the boundaries for periods below 5 days, which is the case of WASP-156b. Understanding the differences between WASP-156b on one side and  WASP-151b and WASP-153b on the other side might allow to shed light on the mechanism responsible for the upper boundary of the Neptunian desert. \citet{2016A&A...589A..75M} proposed two explanations for the origin of the upper boundary of the desert:

\begin{itemize}
\item Gaseous planets can't exist below the upper boundary, because they would lose their gaseous envelope due to stellar insolation \citep[e.g.][]{2014ApJ...792....1L} or Roche-lobe overflow \citep[e.g.][]{2014ApJ...783...54K}.
\item Gaseous planets are formed further away from their parent star and can't migrate below the upper boundary, because at this distance from the star the disk is not dense enough to sustain inward migration.
\end{itemize}
While a detailed analysis of the origin of the Neptunian desert is beyond the scope of this paper, it is still interesting to look into the similarities and differences between WASP-156b and WASP-151b/WASP-153b since they might provide useful hints on the nature of this desert. These three planets  possess similar orbital parameters (see \tab{syspar}). Their ages are subject to caution (as discussed in \sect{agediscrepancy}), but a given estimator provides similar ages for these three stars. Their gyrochronological ages indicate relatively young systems ($\sim 1$\,Gyr for WASP-151 and WASP-153 and $\sim 0.5$\,Gyr for WASP-156), while their isochronal ages indicate $\sim 5\,\mathrm{Gyr}$ old systems. However, their radiative environments are significantly different. WASP-151b and WASP-153b receive a higher bolometric irradiation ($460, 1400$ and $150\,F_{i, \oplus}$ for WASP-151b, WASP153b and WASP-156b respectively). Moreover the spectral type of their host stars are different (early G for WASP-151 and WASP-153 and early K for WASP-156) implying a different spectral content of the irradiation,  especially in extreme ultra-violet (EUV). The EUV flux is particularly interesting in this context since it is the main contributor for exoplanet atmosphere evaporation. \citet[e.g.][]{2007A&A...461.1185L} provided estimates for the EUV flux emitted by stars of different spectral types. According to these estimates, the EUV flux received by WASP-156b is $\sim 3$ times higher than the one received by WASP-151b and WASP-153b, $F_{EUV@1AU}$ is $15\,\mathrm{erg.cm^{-2}.s^{-1}}$ for K type stars and  $5\,\mathrm{erg.cm^{-2}.s^{-1}}$ for G type stars. This suggests that photo-evaporation is the mechanism responsible for the presence of WASP-156b below the upper boundary of the short-period Neptunian desert. WASP-156 may be in the process of losing its gaseous envelope in a short-lived evolutionary phase which places it within the underpopulated short-period Neptunian desert. 

Finally, in the context of the hypothesis formulated in \sect{agediscrepancy}, it is also interesting to mention the alternative explanation defended by \citet{2016ApJ...820L...8M} for the origin of the Neptunian desert. The authors present the desert as the result of high-eccentricity migration of planets that arrive in the vicinity of the Roche limit of their host star and suggest that the slopes and positions of the upper and lower boundaries are a direct consequence of the different mass-radius relations for rocky and gaseous planets.

\begin{figure*}[!htb]
    \centering
    \begin{minipage}{0.5\textwidth}
        \centering
        \includegraphics[width=0.9\textwidth]{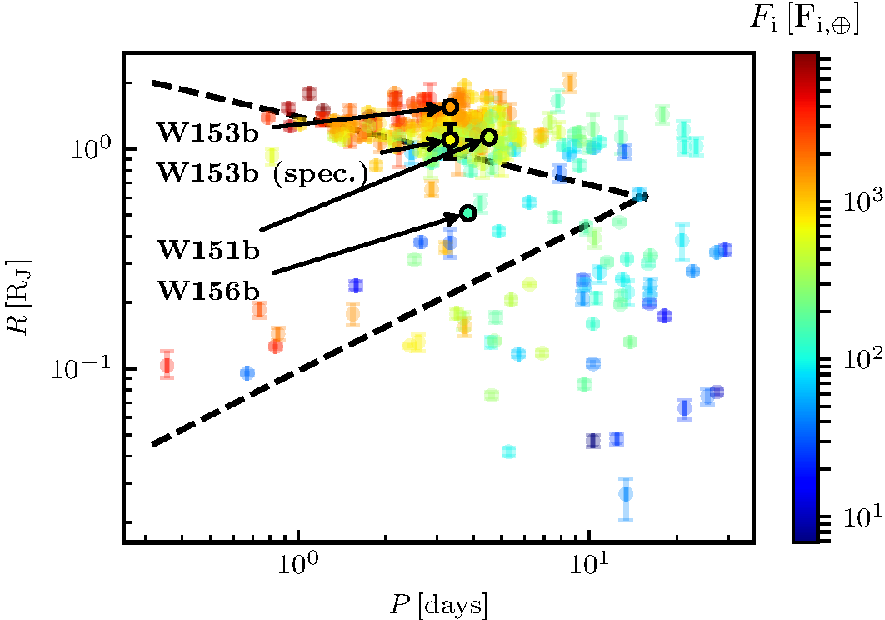}
    \end{minipage}\hfill
    \begin{minipage}{0.5\textwidth}
        \centering
        \includegraphics[width=0.9\textwidth]{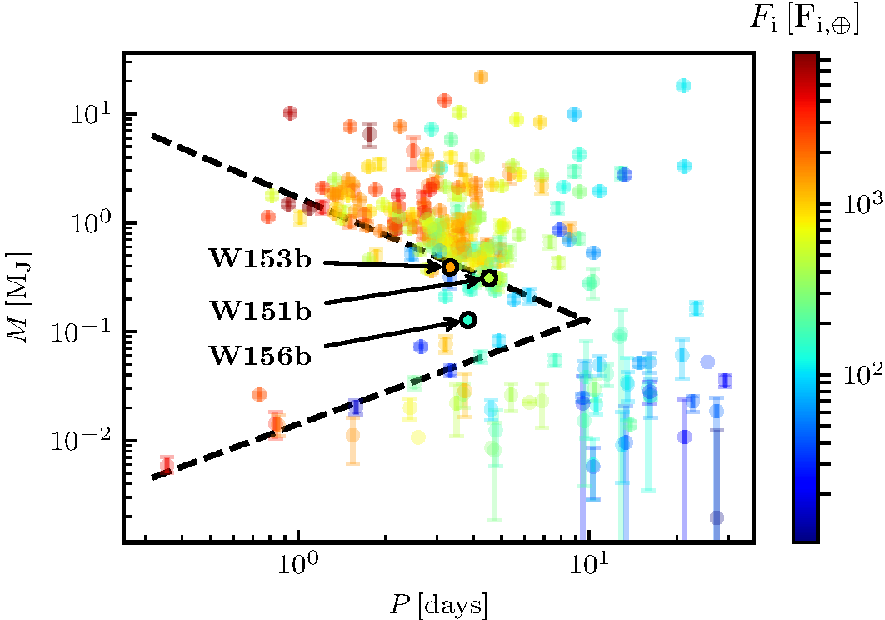}
    \end{minipage}
    \caption{\figlabel{Neptuniandesert}WASP-151b, WASP-153b and WASP-156b in the radius versus orbital period (left) and the mass versus orbital period (right) domains. The colored points correspond to the known exoplanet and the color reflect their bolometric incident flux. WASP-151b, WASP-153b and WASP-156b are circled in black and the black dashed lines correspond to the upper and lower boundaries of the Neptunian desert as reported by \citep{2016A&A...589A..75M}. See \sect{Neptuniandesert} for details.}
\end{figure*}

\begin{acknowledgements}
O.D.S.D acknowledges the support from Fundação para a Ciência e a Tecnologia (FCT) through national funds and by FEDER through COMPETE2020 by grants UID/FIS/04434/2013\&POCI-01-0145-FEDER-007672 and PTDC/FIS-AST/1526/2014\&POCI-01-0145-FEDER-016886. 
SCCB also acknowledges support from FCT through Investigador FCT contract IF/01312/2014/CP1215/CT0004. 
CAH was supported by STFC under grant ST/P000584/1.
A.S.B. acknowledges funding from the European Union Seventh Framework programme (FP7/2007-2013) under grant agreement No. 313014 (ETAEARTH).
D.J.A.B acknowledges funding from the UKSA and the University of Warwick.
The SuperWASP Consortium consists of astronomers primarily from University of Warwick, Queens University Belfast, St Andrews, Keele, Leicester, The Open University, Isaac Newton Group La Palma and Instituto de Astrof\'isica de Canarias. The SuperWASP-N camera is hosted by the Issac Newton Group on La Palma and WASPSouth is hosted by SAAO. We are grateful for their support and assistance. Funding for WASP comes from consortium universities and from the UK's Science and Technology Facilities Council. The research leading to these results has received funding from the European Community’s Seventh Framework Programmes (FP7/2007-2013 and FP7/2013-2016) under grant agreement number RG226604 and 312430 (OPTICON), respectively. 
This work is based on radial velocity observations made at Observatoire de Haute Provence (CNRS), France and at ESO La Silla in (Chile) with the CORALIE Echelle spectrograph mounted on the Swiss telescope. We thank the staff at Haute-Provence Observatory.
TRAPPIST is funded by the Belgian Fund for Scientific Research (Fond National de la Recherche Scientifique, FNRS) under the grant FRFC 2.5.594.09.F, with the participation of the Swiss National Science Fundation (SNF). 
The Swiss {\it Euler} Telescope is a project financed by the Swiss National Science Foundation.
The Liverpool Telescope is operated on the island of La Palma by Liverpool John Moores University in the Spanish Observatorio del Roque de los Muchachos of the Instituto de Astrofisica de Canarias with financial support from the UK Science and Technology Facilities Council.
The Aristarchos telescope is operated on Helmos Observatory by the Institute for Astronomy, Astrophysics, Space Applications and Remote Sensing of the National Observatory of Athens.
This paper includes data collected by the K2 mission. Funding for the K2 mission is provided by the NASA Science Mission directorate.
Used Simbad, Vizier, exoplanet.eu.
The authors also want to thank Pedro Figueira, Nuno Santos and Mahmoud Oshagh for fruitful discussions and Joao Faria for his Python figure styler.
Most of the analyses presented in this paper were performed using the Python language (version 3.5) available at \url{http://www.python.org} and several scientific packages: Numpy \& Scipy \citep{2011CSE1102.1523V}, Pandas \citep{2010PPSC51.56}, Ipython \citep{2007CSE10.1109}, Astropy \citep{2013A&A...558A..33A} and Matplotlib \citep{2007CSE.....9...90H}.

\end{acknowledgements}


\begin{appendix}

\section{Radial velocity measurements}\sectlabel{app:rvmeas}

\begin{table}[ht]
  \centering 
  \caption{\tablabel{rvobsWASP151}Radial velocities of WASP$-$151.}
  \begin{tabular}{cccc}
    \hline
    \hline
    BJD & RV & $\pm$$1\,\sigma$ \\
    -2\,450\,000 & (km.s$^{-1}$) & (km.s$^{-1}$) \\
    \hline
      \multicolumn{3}{c}{SOPHIE$@$OHP} \\
    \hline
6948.4518    & $-$12.420    & 0.011\\
6950.4111     & $-$12.361    & 0.011\\
6974.3650    & $-$12.357    & 0.015\\
6975.3435    & $-$12.399    & 0.011\\
6977.3796    & $-$12.363    & 0.012\\
6979.3954    & $-$12.384    & 0.012\\
6980.3643    & $-$12.417    & 0.011\\
6981.3799    & $-$12.387    & 0.011\\
6982.3904    & $-$12.331    & 0.011\\
7217.6072    & $-$12.335    & 0.016\\
7218.5906    & $-$12.331    & 0.012\\
7221.5984    & $-$12.383    & 0.015\\
7222.6099    & $-$12.305    & 0.017\\
7223.5609    & $-$12.320    & 0.012\\
7224.5609    & $-$12.384    & 0.012\\
7225.6049    & $-$12.391    & 0.012\\
7241.5706    & $-$12.340    & 0.012\\
7242.5667    & $-$12.370    & 0.015\\
7246.5986    & $-$12.345    & 0.012\\
7247.5554    & $-$12.401    & 0.017\\
7275.4448    & $-$12.398    & 0.014\\
7303.4575    & $-$12.398    & 0.013\\
7304.4389    & $-$12.340    & 0.011\\
7306.4954    & $-$12.370    & 0.012\\
7330.3192    & $-$12.405    & 0.015\\
7332.4111     & $-$12.365    & 0.012\\
7333.3746    & $-$12.372    & 0.011\\
7335.3457    & $-$12.345    & 0.011\\
7365.3564    & $-$12.398    & 0.016\\
    
    \hline 
    \multicolumn{3}{c}{CORALIE$@$Euler} \\
    \hline
6864.7421   & $-$12.304    & 0.029\\
7188.9240   & $-$12.318    & 0.047\\
7203.8492   & $-$12.297    & 0.046\\
7270.6872   & $-$12.350    & 0.046\\
7277.6367   & $-$12.245    & 0.034\\
7558.9041   & $-$12.296    & 0.022\\
7584.9040   & $-$12.274    & 0.023\\

    \hline
  \end{tabular}
\end{table}

\begin{table}[ht]
  \centering 
  \caption{\tablabel{rvobsWASP153}Radial velocities of WASP$-$153.}
  \begin{tabular}{cccc}
    \hline
    \hline
    BJD & RV & $\pm$$1\,\sigma$ \\
    -2\,450\,000 & (km.s$^{-1}$) & (km.s$^{-1}$) \\
    \hline
      \multicolumn{3}{c}{SOPHIE$@$OHP} \\
    \hline
6814.4193   & $-$ 28.957    &    0.012\\
6869.3904   & $-$ 29.040    &    0.011\\
6897.4652   & $-$ 28.958    &    0.010\\
6899.4175   & $-$ 29.031    &    0.008\\
6900.3642   & $-$ 28.996    &    0.010\\
6901.4333   & $-$ 29.003    &    0.013\\
6902.4896   & $-$ 29.038    &    0.012\\
6921.3201   & $-$ 28.974    &    0.010\\
6922.3427   & $-$ 29.068    &    0.010\\
6932.3452   & $-$ 29.041    &    0.010\\
6933.3662   & $-$ 28.994    &    0.013\\
6934.3121   & $-$ 28.968    &    0.012\\
6935.3207   & $-$ 29.013    &    0.012\\
6936.3639   & $-$ 29.039    &    0.015\\
6950.3422   & $-$ 29.001    &    0.010\\
6974.3286   & $-$ 28.949    &    0.012\\
6975.2421   & $-$ 29.043    &    0.012\\
6977.2904   & $-$ 28.950    &    0.012\\
6979.2833   & $-$ 29.046    &    0.011\\
6980.2625   & $-$ 28.988    &    0.010\\
6981.2993   & $-$ 28.986    &    0.010\\
6982.3102   & $-$ 29.053    &    0.011\\
7076.7066   & $-$ 29.003    &    0.013\\
7126.5636   & $-$ 29.022    &    0.010\\
7130.5415   & $-$ 28.955    &    0.012\\
7134.5347   & $-$ 28.993    &    0.010\\
7136.6256   & $-$ 29.005    &    0.011\\
7155.4794   & $-$ 29.054    &    0.011\\
7189.3932   & $-$ 29.059    &    0.011\\
7190.4659   & $-$ 28.968    &    0.011\\
7193.4950   & $-$ 28.993    &    0.011\\
7195.5351   & $-$ 29.050    &    0.010\\
7210.5381   & $-$ 28.967    &    0.015\\
7213.3806   & $-$ 29.015    &    0.011\\
7215.5279   & $-$ 29.047    &    0.011\\
7219.5213   & $-$ 29.028    &    0.011\\
7220.4709   & $-$ 28.958    &    0.009\\
7222.5478   & $-$ 29.052    &    0.011\\
7223.3999   & $-$ 28.993    &    0.011\\
7224.3858   & $-$ 28.949    &    0.010\\
7225.4078   & $-$ 29.023    &    0.010\\
7242.4107   & $-$ 29.023    &    0.010\\
7245.4448   & $-$ 29.023    &    0.011\\
7246.3539   & $-$ 29.034    &    0.010\\
7247.3698   & $-$ 28.946    &    0.011\\
7276.3470   & $-$ 29.019    &    0.011\\
7277.3825   & $-$ 28.949    &    0.011\\
7305.2940   & $-$ 29.038    &    0.011\\
7307.2832   & $-$ 28.964    &    0.006\\
7332.2717   & $-$ 29.034    &    0.010\\
7335.3076   & $-$ 29.017    &    0.011\\
7359.2296   & $-$ 29.069    &    0.011\\

    \hline
  \end{tabular}
\end{table}

\begin{table}[ht]
  \centering 
  \caption{\tablabel{rvobsWASP156}Radial velocities of WASP$-$156.}
  \begin{tabular}{cccc}
    \hline
    \hline
    BJD & RV & $\pm$$1\,\sigma$ \\
    -2\,450\,000 & (km.s$^{-1}$) & (km.s$^{-1}$) \\
    \hline
      \multicolumn{3}{c}{SOPHIE$@$OHP} \\
    \hline
6949.5554   &  9.568   &  0.007\\
6950.5013   &  9.584   &  0.007\\
6974.4265   &  9.601   &  0.008\\
6975.4051   &  9.601   &  0.007\\
6977.4038   &  9.585   &  0.007\\
6980.4032   &  9.573   &  0.007\\
6981.4348   &  9.599   &  0.007\\
6982.4510   &  9.606   &  0.007\\
7018.3516   &  9.592   &  0.008\\
7046.2604   &  9.562   &  0.007\\
7241.6241   &  9.571   &  0.008\\
7242.6182   &  9.610   &  0.008\\
7246.6178   &  9.604   &  0.007\\
7247.5997   &  9.611   &  0.007\\
7275.5475   &  9.562   &  0.007\\
7276.5779   &  9.590   &  0.007\\
7303.5231   &  9.593   &  0.005\\
7304.4994   &  9.602   &  0.005\\
7306.5282   &  9.560   &  0.007\\
7331.5277   &  9.614   &  0.005\\
7333.4191   &  9.575   &  0.005\\
7334.5369   &  9.589   &  0.005\\
7335.3780   &  9.590   &  0.007\\
7364.4221   &  9.557   &  0.007\\
7365.3417   &  9.594   &  0.008\\
7396.2886   &  9.601   &  0.006\\
7398.3562   &  9.574   &  0.005\\
7400.2839   &  9.612   &  0.009\\
7402.3068   &  9.567   &  0.005\\
7607.6081   &  9.623   &  0.005\\
7623.5766   &  9.608   &  0.006\\
7624.6315   &  9.573   &  0.005\\
7625.5790   &  9.581   &  0.004\\
7626.5939   &  9.606   &  0.005\\
7627.6071   &  9.594   &  0.005\\
7628.5911   &  9.577   &  0.005\\
7659.5871   &  9.575   &  0.005\\
7660.5332   &  9.605   &  0.005\\
7661.5583   &  9.609   &  0.005\\
7682.6195   &  9.581   &  0.006\\
7719.4024   &  9.607   &  0.005\\
7744.3552   &  9.571   &  0.004\\
7745.3334   &  9.589   &  0.005\\
7746.3460   &  9.598   &  0.004\\
    
    \hline 
    \multicolumn{3}{c}{CORALIE$@$Euler} \\
    \hline
6920.8939   &  9.638   &  0.011\\
6922.8367   &  9.613   &  0.009\\
6924.7687   &  9.657   &  0.006\\
7016.6163   &  9.645   &  0.009\\
7039.5437   &  9.620   &  0.014\\
7041.5431   &  9.611   &  0.015\\
7261.8518   &  9.645   &  0.012\\
7339.6856   &  9.636   &  0.013\\
7417.5361   &  9.618   &  0.007\\
7587.8980   &  9.637   &  0.009\\
7681.6026   &  9.624   &  0.008\\
7692.6865   &  9.651   &  0.009\\
7752.6070   &  9.628   &  0.009\\

    \hline
  \end{tabular}
\end{table}

\begin{table*}
\begin{center}
\caption{\tablabel{priorandposteriors}\textit{Prior} functions for each free parameters.}
\begin{tabular}{lccc} 
\textit{Parameters} & WASP-151b & WASP-153b & WASP-156b \\
\hline \\[-6pt]

${P}$\ [days]					& $\mathcal{N}(4.5334,\,0.003)$		& $\mathcal{N}(3.333,\,0.001)$	& $\mathcal{N}(3.83616,\,3\,10^{5})$ \\[3pt]
${t_c}$\ [BJD - 2\,400\,000]	& $\mathcal{N}(57741.0,\,0.1)$		& $\mathcal{N}(53142.5,\,0.1)[\mu - 2\sigma,\,\mu + 2\sigma]$	& $\mathcal{N}(54677.71,\,0.01)$ \\[3pt]
${\sqrt{e}\cos \omega_*}$		& \multicolumn{3}{c}{-------------\ $\mathcal{N}(0, 0.05)[0,\,1/\sqrt{2}]$\ -------------} \\[3pt]
${\sqrt{e}\sin \omega_*}$		& \multicolumn{3}{c}{-------------\ $\mathcal{N}(0, 0.05)[0,\,1/\sqrt{2}]$\ -------------} \\[3pt]
${\cos i}$						& \multicolumn{3}{c}{-------------\ $\mathcal{N}(0, 0.1)[0.,\,1.]$\ -------------} \\[3pt]
${a / R_*}$						& $\mathcal{N}(8.67, 1)[1.,\,30]$	& $\mathcal{N}(8.4,\,1)[1,\,50]$	& $\mathcal{U}(1.,\,30)$ \\[3pt]
${R_p / R_*}$ [\%]				& \multicolumn{3}{c}{-------------\ $\mathcal{U}(1, 20)$\ -------------} \\[3pt]
${K}$ [\ms]						& \multicolumn{3}{c}{-------------\ $\mathcal{U}(0, 1)$\ -------------} \\[3pt]
${v0_{\textrm{SOPHIE}}}$ [\kms]	& $\mathcal{N}(-12.4,\,0.02)$				& $\mathcal{N}(-29,\,0.1)$		& $\mathcal{N}(9.58,\,0.01)$ \\[3pt]
${u_{\textrm{Johson}\,R}}$		& $\mathcal{N}(0.4781,\,0.0022)$				& $\mathcal{N}(0.4859,\,0.0020)$ 		& $\mathcal{N}(0.5911,\,0.0032)$ \\[3pt]
${v_{\textrm{Johson}\,R}}$		& $\mathcal{N}(0.1304,\,0.0055)$				& $\mathcal{N}(0.1258,\,0.0053)$ 		& $\mathcal{N}(0.0805,\,0.0082)$ \\[3pt]
${u_{\textrm{Sloan}\,z}}$		& $\mathcal{N}(0.3412,\,0.0013)$		& 							&\\[3pt]
${v_{\textrm{Sloan}\,z}}$		& $\mathcal{N}(0.1269,\,0.0039)$		& 							&\\[3pt]
${u_{\textrm{Kp}}}$				& $\mathcal{N}(0.5501,\,0.0025)$		& 							&\\[3pt]
${v_{\textrm{Kp}}}$				& $\mathcal{N}(0.1191,\,0.0054)$		& 							&\\[3pt]
${u_{\textrm{NGTS}}}$			& $\mathcal{N}(0.4861,\,0.0021)$		& 							&\\[3pt]
${v_{\textrm{NGTS}}}$			& $\mathcal{N}(0.1286,\,0.0052)$				& 							&\\[3pt]
${u_{\textrm{Johson}\,I}}$		& 									&  							& $\mathcal{N}(0.4608,\,0.0024)$ \\[3pt]
${v_{\textrm{Johson}\,I}}$		& 									&  							& $\mathcal{N}(0.1013,\,0.0070)$ \\[3pt]
${u_{\textrm{Gunn}\,r}}$		& 									&  		&					$\mathcal{N}(0.4608,\,0.0024)$ \\[3pt]
${v_{\textrm{Gunn}\,r}}$		& 									&  		&					$\mathcal{N}(0.0812,\,0.0085)$ \\[3pt]
${\Delta\textrm{RV}_{\textrm{CORALIE/SOPHIE}}}$ [\kms]	& $\mathcal{N}(0.05,\,0.01)$							& 						& $\mathcal{N}(0.05,\,0.005)$\\[3pt]
${\ln f_{\sigma \textrm{SOPHIE}}}$							& \multicolumn{3}{c}{-------------\ $\mathcal{N}(0,\,0.05)$\ -------------} \\[3pt]
${\ln f_{\sigma \textrm{CORALIE}}}$							& $\mathcal{N}(0,\,0.05)$							& 						& $\mathcal{N}(0,\,0.05)$ \\[3pt]
${\ln f_{\sigma \textrm{WASP}}}$							& \multicolumn{3}{c}{-------------\ $\mathcal{N}(0,\,0.05)$\ -------------} \\[3pt]
${\ln f_{\sigma \textrm{K2}}}$								& $\mathcal{N}(0,\,0.05)$							&  						&\\[3pt]
${\ln f_{\sigma \textrm{EulerCam}}}$							& $\mathcal{N}(0,\,0.05)$	&	& $\mathcal{N}(0,\,0.05)$ \\[3pt]
${\ln f_{\sigma \textrm{TRAPPIST}}}$							& $\mathcal{N}(0,\,0.05)$							&  						&\\[3pt]
$\ln f_{\sigma \textrm{Liverpool}}$							&	& $\mathcal{N}(0,\,0.05)$ & \\[3pt]
${\ln f_{\sigma \textrm{RISE2}}}$							&	& $\mathcal{N}(0,\,0.05)$ & \\[3pt]
${\ln f_{\sigma \textrm{NITES}}}$								& 											&  						& $\mathcal{N}(0,\,0.05)$\\[3pt]
${\ln f_{\sigma \textrm{IAC80}}}$				& $\mathcal{N}(0,\,0.05)$						&  						& \\[3pt]
$\Delta F_{\textsc{oot}, \textrm{IAC80}}$		& $\mathcal{N}(0.005,\,0.01)$	&	&\\[3pt]
															& $\mathcal{N}(0.0,\,0.01)$	&	& \\[3pt]
															& $\mathcal{N}(0.0,\,0.01)$	&  						&\\[3pt]
$\Delta F_{\textsc{oot}, \textrm{IAC80}}^\prime$ [day$^{-1}$]	& $\mathcal{N}(0.0,\,0.33)$	&	&\\[3pt]
															& $\mathcal{N}(0.0,\,0.1)$	&	& \\[3pt]
															& $\mathcal{N}(0.0,\,0.1)$	&	& \\[3pt]
$\Delta F_{\textsc{oot}, \textrm{TRAPPIST}}$				& $\mathcal{N}(0.0,\,0.001)$	&	& \\[3pt]	
$\Delta F_{\textsc{oot}, \textrm{TRAPPIST}}^\prime$	 [day$^{-1}$]	& $\mathcal{N}(0.0,\,0.005)$	&	&\\[3pt]
$\Delta F_{\textsc{oot}, \textrm{EulerCam}}$						&	&	& $\mathcal{N}(-0.00036,\,0.001)$ \\[3pt]	
$\Delta F_{\textsc{oot}, \textrm{EulerCam}}^\prime$ [day$^{-1}$]	&	&	& $\mathcal{N}(0.0012,\,0.001)$ \\[3pt]
$\Delta F_{\textsc{oot}, \textrm{EulerCam}}^{\prime\prime}$ [day$^{-2}$]	&	&	& $\mathcal{N}(-0.001,\,0.001)$ \\[3pt]
\hline
\end{tabular}
\end{center}
\textbf{Notes:} $\mathcal{N}(\mu,\,\sigma)$ designate normal distributions of mean $\mu$ and standard deviation $\sigma$.\\
$\mathcal{N}(\mu,\,\sigma)[\mathrm{min},\,\mathrm{max}]$ designate truncated normal distributions with min and max as minimum and maximum value.\\ 
$\mathcal{U}(\mathrm{min},\,\mathrm{max})$ designate uniform distributions with min and max as minimum and maximum value.\\
-------------\ \textit{distribution}\ ------------- indicates that the same \textit{prior} distribution has been used for the analysis of the three systems.\\ 
\end{table*}

\begin{table*}
 \caption{\tablabel{fullbagemassoutput}Complete \texttt{bagemass} output table giving the Bayesian mass and age estimates for WASP-151, WASP-153 and WASP-156. Columns 2, 3 and 4 give the maximum-likelihood estimates of the age, mass, and initial metallicity, respectively. Column 5 is the chi-squared statistic of the fit for the parameter values in columns 2, 3, and 4.  Columns 6 and 7 give the mean and standard deviation of their marginalized \textit{posterior} distributions. Column 8 ($p_{\rm MS}$) is the probability that the star is
still on the main sequence. The systematic errors on the mass and age due to uncertainties in the mixing length and helium abundance are given in columns 9 to 12. For more details see \sect{stelmod} and \citet{2015A&A...575A..36M}.
}
\begin{tabular}{@{}lccccccccccc}
\hline
Star &
  \multicolumn{1}{c}{$\tau_{\rm iso, b}$ [Gyr]} &
  \multicolumn{1}{c}{$M_{\rm b}$[$M_{\odot}$]} &
  \multicolumn{1}{c}{$\mathrm{[Fe/H]}_\mathrm{i, b}$} &
  \multicolumn{1}{c}{$\chi^2$}&
  \multicolumn{1}{c}{$\langle \tau_{\rm iso} \rangle$ [Gyr]}  &
  \multicolumn{1}{c}{$\langle M_{\star} \rangle$ [$M_{\odot}$]} &
  \multicolumn{1}{c}{$p_{\rm MS}$} &
  \multicolumn{1}{c}{$\sigma_{\tau, Y}$}  &
  \multicolumn{1}{c}{$\sigma_{\tau,\alpha}$} &
  \multicolumn{1}{c}{$\sigma_{M, Y}$}  &
  \multicolumn{1}{c}{$\sigma_{M,\alpha}$}  \\
\hline
 \noalign{\smallskip}
WASP-151         &   5.0 &   1.07 &$ +0.162 $&  0.002 &$ 5.13 \pm  1.33 $&$  1.077 \pm 0.048 $& $  1.00 $&$ -0.42 $&$  2.59 $&$ 0.044 $&$ -0.049 $ \\
WASP-153         &$ 3.8 $&$ 1.35 $&$ +0.339 $&$ 0.015 $&$ 4.00 \pm  0.77 $&$ 1.336 \pm 0.065 $&$  0.87 $&$ -0.08 $&$ 0.79 $&$ 0.052 $&$ -0.021 $ \\
WASP-156         &   0.5 &   0.87 &$ +0.255 $&  0.02 &$ 6.50 \pm  4.03 $&$  0.842 \pm 0.036 $&$  1.00 $&$  -0.68 $&$  1.37 $&$ 0.034 $&$ -0.017 $ \\
 \noalign{\smallskip}
\hline
\end{tabular}   
\end{table*}     

\end{appendix}

\end{document}